\documentclass[fleqn,usenatbib]{rasti}

\usepackage{newtxtext,newtxmath}
\usepackage{anyfontsize}

\usepackage[T1]{fontenc}

\DeclareRobustCommand{\VAN}[3]{#2}
\let\VANthebibliography\thebibliography
\def\thebibliography{\DeclareRobustCommand{\VAN}[3]{##3}\VANthebibliography}

\newcommand{\FITNAperiod}{\mbox{$3.7395671$}}
\newcommand{\FITNArb}{\mbox{$0.1311$}}
\newcommand{\FITNArsuma}{\mbox{$0.102827$}}
\newcommand{\FITNAtc}{\mbox{$2460505.6096463003$}}
\newcommand{\FITNAqone}{\mbox{$0.387$}}
\newcommand{\FITNAqtwo}{\mbox{$0.408$}}
\newcommand{\FITNAinc}{\mbox{$0.075327$}}

\newcommand{\CMOSNArb}{\mbox{$0.1307_{-0.0012}^{+0.0011}$}}
\newcommand{\CMOSNArsuma}{\mbox{$0.1058\pm0.0034$}}%
\newcommand{\CMOSNAinc}{\mbox{$0.0784\pm0.0032$}}
\newcommand{\CMOSNAtc}{\mbox{$2460505.61003\pm0.00040$}}
\newcommand{\CMOSNAperiod}{\mbox{$3.7395672_{-0.0000013}^{+0.0000012}$}}
\newcommand{\CMOSNAqone}{\mbox{$0.389\pm0.015$}}
\newcommand{\CMOSNAqtwo}{\mbox{$0.447\pm0.093$}}

\newcommand{\CCDNArb}{\mbox{$0.1310\pm0.0011$}}
\newcommand{\CCDNArsuma}{\mbox{$0.1026\pm0.0034$}}%
\newcommand{\CCDNAinc}{\mbox{$0.0734_{-0.0037}^{+0.0035}$}}
\newcommand{\CCDNAtc}{\mbox{$2460505.60988\pm0.00040$}}
\newcommand{\CCDNAperiod}{\mbox{$3.7395670\pm0.0000013$}}
\newcommand{\CCDNAqone}{\mbox{$0.388_{-0.016}^{+0.014}$}}
\newcommand{\CCDNAqtwo}{\mbox{$0.400\pm0.097$}}

\newcommand{\FITNBperiod}{\mbox{$2.1846730$}}
\newcommand{\FITNBrb}{\mbox{$0.1025$}}
\newcommand{\FITNBrsuma}{\mbox{$0.172266$}}
\newcommand{\FITNBtc}{\mbox{$2458553.711158$}}
\newcommand{\FITNBqone}{\mbox{$0.357$}}
\newcommand{\FITNBqtwo}{\mbox{$0.393$}}
\newcommand{\FITNBinc}{\mbox{$0.027922$}}

\newcommand{\CMOSNBrb}{\mbox{$0.10231\pm0.00055$}}
\newcommand{\CMOSNBrsuma}{\mbox{$0.1771_{-0.0026}^{+0.0050}$}}%
\newcommand{\CMOSNBinc}{\mbox{$0.019_{-0.039}^{+0.028}$}}
\newcommand{\CMOSNBtc}{\mbox{$2460506.80136\pm0.00070$}}
\newcommand{\CMOSNBperiod}{\mbox{$2.18467329_{-0.00000090}^{+0.0000013}$}}
\newcommand{\CMOSNBqone}{\mbox{$0.359_{-0.015}^{+0.012}$}}
\newcommand{\CMOSNBqtwo}{\mbox{$0.366_{-0.065}^{+0.060}$}}

\newcommand{\CCDNBrb}{\mbox{$0.10250_{-0.00053}^{+0.00047}$}}
\newcommand{\CCDNBrsuma}{\mbox{$0.1732_{-0.0048}^{+0.0057}$}}%
\newcommand{\CCDNBinc}{\mbox{$0.048_{-0.036}^{+0.017}$}}
\newcommand{\CCDNBtc}{\mbox{$2460506.80373\pm0.00073$}}
\newcommand{\CCDNBperiod}{\mbox{$2.1846729_{-0.0000013}^{+0.0000014}$}}
\newcommand{\CCDNBqone}{\mbox{$0.359\pm0.011$}}
\newcommand{\CCDNBqtwo}{\mbox{$0.369\pm0.079$}}

\newcommand{\FITNCperiod}{\mbox{$0.672474140$}}
\newcommand{\FITNCrb}{\mbox{$0.08155$}}
\newcommand{\FITNCrsuma}{\mbox{$0.476874$}}
\newcommand{\FITNCtc}{\mbox{$2459378.455370$}}
\newcommand{\FITNCqone}{\mbox{$0.305$}}
\newcommand{\FITNCqtwo}{\mbox{$0.375$}}
\newcommand{\FITNCinc}{\mbox{$0.329855$}}

\newcommand{\CMOSNCrb}{\mbox{$0.08155\pm0.00022$}}
\newcommand{\CMOSNCrsuma}{\mbox{$0.4701\pm0.0073$}}%
\newcommand{\CMOSNCinc}{\mbox{$0.3297\pm0.0060$}}
\newcommand{\CMOSNCtc}{\mbox{$2460509.556868\pm0.000074$}}
\newcommand{\CMOSNCperiod}{\mbox{$0.672474141\pm0.000000028$}}
\newcommand{\CMOSNCqone}{\mbox{$0.304\pm0.008$}}
\newcommand{\CMOSNCqtwo}{\mbox{$0.379\pm0.074$}}

\newcommand{\CCDNCrb}{\mbox{$0.08155\pm0.00022$}}
\newcommand{\CCDNCrsuma}{\mbox{$0.4649_{-0.0076}^{+0.0080}$}}%
\newcommand{\CCDNCinc}{\mbox{$0.3299\pm0.0059$}}
\newcommand{\CCDNCtc}{\mbox{$2460509.556874\pm0.000074$}}
\newcommand{\CCDNCperiod}{\mbox{$0.672474141_{-0.000000029}^{+0.000000027}$}}
\newcommand{\CCDNCqone}{\mbox{$0.305\pm0.008$}}
\newcommand{\CCDNCqtwo}{\mbox{$0.377\pm0.074$}}

\newcommand{\FITNDperiod}{\mbox{$4.1662541$}}
\newcommand{\FITNDrb}{\mbox{$0.1190$}}
\newcommand{\FITNDrsuma}{\mbox{$0.119807$}}
\newcommand{\FITNDtc}{\mbox{$2457612.49947$}}
\newcommand{\FITNDqone}{\mbox{$0.350$}}
\newcommand{\FITNDqtwo}{\mbox{$0.385$}}
\newcommand{\FITNDinc}{\mbox{$0.024258$}}

\newcommand{\CMOSNDrb}{\mbox{$0.11793\pm0.0010$}}
\newcommand{\CMOSNDrsuma}{\mbox{$0.1212_{-0.0025}^{+0.0028}$}}%
\newcommand{\CMOSNDinc}{\mbox{$0.021_{-0.013}^{+0.011}$}}
\newcommand{\CMOSNDtc}{\mbox{$2460524.7109\pm0.0011$}}
\newcommand{\CMOSNDperiod}{\mbox{$4.1662541\pm0.0000016$}}
\newcommand{\CMOSNDqone}{\mbox{$0.349\pm0.010$}}
\newcommand{\CMOSNDqtwo}{\mbox{$0.341\pm0.069$}}

\newcommand{\CCDNDrb}{\mbox{$0.12013\pm0.0010$}}
\newcommand{\CCDNDrsuma}{\mbox{$0.1205_{-0.0023}^{+0.0025}$}}%
\newcommand{\CCDNDinc}{\mbox{$0.017_{-0.012}^{+0.010}$}}
\newcommand{\CCDNDtc}{\mbox{$2460524.7110_{-0.0011}^{+0.0010}$}}
\newcommand{\CCDNDperiod}{\mbox{$4.1662541\pm0.0000016$}}
\newcommand{\CCDNDqone}{\mbox{$0.351\pm0.010$}}
\newcommand{\CCDNDqtwo}{\mbox{$0.416\pm0.068$}}

\newcommand{\FITNEperiod}{\mbox{$2.07275965$}}
\newcommand{\FITNErb}{\mbox{$0.1091$}}
\newcommand{\FITNErsuma}{\mbox{$0.168300$}}
\newcommand{\FITNEtc}{\mbox{$2458554.475352$}}
\newcommand{\FITNEqone}{\mbox{$0.369$}}
\newcommand{\FITNEqtwo}{\mbox{$0.400$}}
\newcommand{\FITNEinc}{\mbox{$0.034899$}}

\newcommand{\CMOSNErb}{\mbox{$0.10856\pm0.00077$}}
\newcommand{\CMOSNErsuma}{\mbox{$0.1686\pm0.0035$}}%
\newcommand{\CMOSNEinc}{\mbox{$0.046_{-0.013}^{+0.010}$}}
\newcommand{\CMOSNEtc}{\mbox{$2460527.74264\pm0.00020$}}
\newcommand{\CMOSNEperiod}{\mbox{$2.07275967_{-0.00000021}^{+0.00000020}$}}
\newcommand{\CMOSNEqone}{\mbox{$0.369\pm0.009$}}
\newcommand{\CMOSNEqtwo}{\mbox{$0.410\pm0.068$}}

\newcommand{\CCDNErb}{\mbox{$0.10913\pm0.00079$}}
\newcommand{\CCDNErsuma}{\mbox{$0.1699_{-0.0026}^{+0.0028}$}}%
\newcommand{\CCDNEinc}{\mbox{$0.023_{-0.017}^{+0.014}$}}
\newcommand{\CCDNEtc}{\mbox{$2460527.74252\pm0.00021$}}
\newcommand{\CCDNEperiod}{\mbox{$2.07275965\pm0.00000021$}}
\newcommand{\CCDNEqone}{\mbox{$0.369\pm0.009$}}
\newcommand{\CCDNEqtwo}{\mbox{$0.389\pm0.068$}}

\newcommand{\FITNFperiod}{\mbox{$2.07275965$}}
\newcommand{\FITNFrb}{\mbox{$0.1091$}}
\newcommand{\FITNFrsuma}{\mbox{$0.168300$}}
\newcommand{\FITNFtc}{\mbox{$2458554.475352$}}
\newcommand{\FITNFqone}{\mbox{$0.369$}}
\newcommand{\FITNFqtwo}{\mbox{$0.400$}}
\newcommand{\FITNFinc}{\mbox{$0.034899$}}

\newcommand{\CMOSNFrb}{\mbox{$0.10825\pm0.00078$}}
\newcommand{\CMOSNFrsuma}{\mbox{$0.1681_{-0.0031}^{+0.0033}$}}%
\newcommand{\CMOSNFinc}{\mbox{$0.040_{-0.016}^{+0.012}$}}
\newcommand{\CMOSNFtc}{\mbox{$2460529.81532\pm0.00020$}}
\newcommand{\CMOSNFperiod}{\mbox{$2.07275964\pm0.00000021$}}
\newcommand{\CMOSNFqone}{\mbox{$0.369\pm0.009$}}
\newcommand{\CMOSNFqtwo}{\mbox{$0.385\pm0.067$}}

\newcommand{\CCDNFrb}{\mbox{$0.10865\pm0.00079$}}
\newcommand{\CCDNFrsuma}{\mbox{$0.1681\pm0.0035$}}%
\newcommand{\CCDNFinc}{\mbox{$0.045_{-0.014}^{+0.010}$}}
\newcommand{\CCDNFtc}{\mbox{$2460529.81530_{-0.00020}^{+0.00019}$}}
\newcommand{\CCDNFperiod}{\mbox{$2.07275965\pm0.00000021$}}
\newcommand{\CCDNFqone}{\mbox{$0.369\pm0.009$}}
\newcommand{\CCDNFqtwo}{\mbox{$0.380\pm0.070$}}

\newcommand{\FITNGperiod}{\mbox{$2.1846730$}}
\newcommand{\FITNGrb}{\mbox{$0.1025$}}
\newcommand{\FITNGrsuma}{\mbox{$0.172266$}}
\newcommand{\FITNGtc}{\mbox{$2458553.711158$}}
\newcommand{\FITNGqone}{\mbox{$0.357$}}
\newcommand{\FITNGqtwo}{\mbox{$0.393$}}
\newcommand{\FITNGinc}{\mbox{$0.027922$}}

\newcommand{\CMOSNGrb}{\mbox{$0.10251_{-0.00059}^{+0.00036}$}}
\newcommand{\CMOSNGrsuma}{\mbox{$0.1783\pm0.0050$}}%
\newcommand{\CMOSNGinc}{\mbox{$0.031_{-0.090}^{+0.029}$}}
\newcommand{\CMOSNGtc}{\mbox{$2460530.83412_{-0.00050}^{+0.00045}$}}
\newcommand{\CMOSNGperiod}{\mbox{$2.1846731_{-0.0000012}^{+0.0000010}$}}
\newcommand{\CMOSNGqone}{\mbox{$0.354_{-0.009}^{+0.008}$}}
\newcommand{\CMOSNGqtwo}{\mbox{$0.446_{-0.049}^{+0.065}$}}

\newcommand{\CCDNGrb}{\mbox{$0.10215\pm0.00048$}}
\newcommand{\CCDNGrsuma}{\mbox{$0.1765_{-0.0051}^{+0.0055}$}}%
\newcommand{\CCDNGinc}{\mbox{$0.051_{-0.038}^{+0.016}$}}
\newcommand{\CCDNGtc}{\mbox{$2460530.83438\pm0.00056$}}
\newcommand{\CCDNGperiod}{\mbox{$2.1846729\pm0.0000014$}}
\newcommand{\CCDNGqone}{\mbox{$0.359\pm0.012$}}
\newcommand{\CCDNGqtwo}{\mbox{$0.407\pm0.079$}}

\newcommand{\FITNHperiod}{\mbox{$4.1662541$}}
\newcommand{\FITNHrb}{\mbox{$0.1190$}}
\newcommand{\FITNHrsuma}{\mbox{$0.119807$}}
\newcommand{\FITNHtc}{\mbox{$2457612.49947$}}
\newcommand{\FITNHqone}{\mbox{$0.350$}}
\newcommand{\FITNHqtwo}{\mbox{$0.385$}}
\newcommand{\FITNHinc}{\mbox{$0.024258$}}

\newcommand{\CMOSNHrb}{\mbox{$0.11958\pm0.00099$}}
\newcommand{\CMOSNHrsuma}{\mbox{$0.1204_{-0.0025}^{+0.0028}$}}%
\newcommand{\CMOSNHinc}{\mbox{$0.022_{-0.013}^{+0.010}$}}
\newcommand{\CMOSNHtc}{\mbox{$2460549.70892\pm0.00099$}}
\newcommand{\CMOSNHperiod}{\mbox{$4.1662541\pm0.0000016$}}
\newcommand{\CMOSNHqone}{\mbox{$0.350\pm0.010$}}
\newcommand{\CMOSNHqtwo}{\mbox{$0.358\pm0.064$}}

\newcommand{\CCDNHrb}{\mbox{$0.1167_{-0.0013}^{+0.0014}$}}
\newcommand{\CCDNHrsuma}{\mbox{$0.1206_{-0.0031}^{+0.0033}$}}%
\newcommand{\CCDNHinc}{\mbox{$0.0354_{-0.013}^{+0.0090}$}}
\newcommand{\CCDNHtc}{\mbox{$2460549.7083\pm0.0011$}}
\newcommand{\CCDNHperiod}{\mbox{$4.1662541\pm0.0000015$}}
\newcommand{\CCDNHqone}{\mbox{$0.350_{-0.011}^{+0.010}$}}
\newcommand{\CCDNHqtwo}{\mbox{$0.368\pm0.072$}}

\newcommand{\FITNIperiod}{\mbox{$2.8807198$}}
\newcommand{\FITNIrb}{\mbox{$0.11449$}}
\newcommand{\FITNIrsuma}{\mbox{$0.136414$}}
\newcommand{\FITNItc}{\mbox{$2459291.64004$}}
\newcommand{\FITNIqone}{\mbox{$0.372$}}
\newcommand{\FITNIqtwo}{\mbox{$0.400$}}
\newcommand{\FITNIinc}{\mbox{$0.107652$}}

\newcommand{\CMOSNIrb}{\mbox{$0.1154_{-0.0030}^{+0.0032}$}}
\newcommand{\CMOSNIrsuma}{\mbox{$0.1360\pm0.0029$}}%
\newcommand{\CMOSNIinc}{\mbox{$0.1088\pm0.0027$}}
\newcommand{\CMOSNItc}{\mbox{$2460550.51237\pm0.00068$}}
\newcommand{\CMOSNIperiod}{\mbox{$2.8807197_{-0.0000027}^{+0.0000029}$}}
\newcommand{\CMOSNIqone}{\mbox{$0.372\pm0.008$}}
\newcommand{\CMOSNIqtwo}{\mbox{$0.397\pm0.070$}}

\newcommand{\CCDNIrb}{\mbox{$0.1133\pm0.0039$}}
\newcommand{\CCDNIrsuma}{\mbox{$0.1351\pm0.0032$}}%
\newcommand{\CCDNIinc}{\mbox{$0.1107\pm0.0031$}}
\newcommand{\CCDNItc}{\mbox{$2460550.51468\pm0.00086$}}
\newcommand{\CCDNIperiod}{\mbox{$2.8807198\pm0.0000028$}}
\newcommand{\CCDNIqone}{\mbox{$0.372\pm0.008$}}
\newcommand{\CCDNIqtwo}{\mbox{$0.401_{-0.069}^{+0.074}$}}

\newcommand{\FITNKperiod}{\mbox{$4.156736$}}
\newcommand{\FITNKrb}{\mbox{$0.0705$}}
\newcommand{\FITNKrsuma}{\mbox{$0.121078$}}
\newcommand{\FITNKtc}{\mbox{$2460551.738836$}}
\newcommand{\FITNKqone}{\mbox{$0.330$}}
\newcommand{\FITNKqtwo}{\mbox{$0.373$}}
\newcommand{\FITNKinc}{\mbox{$0.007505$}}

\newcommand{\CMOSNKrb}{\mbox{$0.0728\pm0.0010$}}
\newcommand{\CMOSNKrsuma}{\mbox{$0.1250\pm0.0015$}}%
\newcommand{\CMOSNKinc}{\mbox{$0.0042\pm0.0063$}}
\newcommand{\CMOSNKtc}{\mbox{$2460551.7454\pm0.0013$}}
\newcommand{\CMOSNKperiod}{\mbox{$4.1567360\pm0.0000020$}}
\newcommand{\CMOSNKqone}{\mbox{$0.329\pm0.006$}}
\newcommand{\CMOSNKqtwo}{\mbox{$0.372\pm0.065$}}

\newcommand{\CCDNKrb}{\mbox{$0.06924\pm0.0010$}}
\newcommand{\CCDNKrsuma}{\mbox{$0.1233\pm0.0019$}}%
\newcommand{\CCDNKinc}{\mbox{$0.0053\pm0.0076$}}
\newcommand{\CCDNKtc}{\mbox{$2460551.7421\pm0.0026$}}
\newcommand{\CCDNKperiod}{\mbox{$4.1567359\pm0.0000020$}}
\newcommand{\CCDNKqone}{\mbox{$0.329\pm0.006$}}
\newcommand{\CCDNKqtwo}{\mbox{$0.350\pm0.063$}}

\newcommand{\FITNTperiod}{\mbox{$1.338231466$}}
\newcommand{\FITNTrb}{\mbox{$0.15201$}}
\newcommand{\FITNTrsuma}{\mbox{$0.211339$}}
\newcommand{\FITNTtc}{\mbox{$2455804.515752$}}
\newcommand{\FITNTqone}{\mbox{$0.385$}}
\newcommand{\FITNTqtwo}{\mbox{$0.400$}}
\newcommand{\FITNTinc}{\mbox{$0.016405$}}

\newcommand{\CMOSNTrb}{\mbox{$0.15211\pm0.00039$}}
\newcommand{\CMOSNTrsuma}{\mbox{$0.2122\pm0.0016$}}%
\newcommand{\CMOSNTinc}{\mbox{$0.013_{-0.013}^{+0.012}$}}
\newcommand{\CMOSNTtc}{\mbox{$2460525.796300\pm0.000080$}}
\newcommand{\CMOSNTperiod}{\mbox{$1.338231464\pm0.000000023$}}
\newcommand{\CMOSNTqone}{\mbox{$0.386\pm0.012$}}
\newcommand{\CMOSNTqtwo}{\mbox{$0.363_{-0.075}^{+0.069}$}}

\newcommand{\CCDNTrb}{\mbox{$0.15215\pm0.00039$}}
\newcommand{\CCDNTrsuma}{\mbox{$0.2105\pm0.0017$}}%
\newcommand{\CCDNTinc}{\mbox{$0.016_{-0.014}^{+0.013}$}}
\newcommand{\CCDNTtc}{\mbox{$2460525.796306\pm0.000079$}}
\newcommand{\CCDNTperiod}{\mbox{$1.338231467\pm0.000000023$}}
\newcommand{\CCDNTqone}{\mbox{$0.387\pm0.012$}}
\newcommand{\CCDNTqtwo}{\mbox{$0.419\pm0.074$}}

\newcommand{\tess}{{\it TESS}}

\usepackage{graphicx}	
\usepackage{amsmath}	
\usepackage{orcidlink}
\usepackage[normalem]{ulem}
\usepackage{soul}

\usepackage{siunitx} 
\newcommand{\ngts}{{NGTS}}

\newcommand{\eps}{{e$^-$\,pix$^{-1}$\,sec$^{-1}$}}

\title[Precise Photometry]{High-Precision Photometry with a scientific CMOS Camera: II On-Sky Testing of the Marana camera at the NGTS facility}

\author[Ioannis Apergis et al.]{
Ioannis~Apergis$^{\orcidlink{0009-0004-7473-4573}}$,$^{1,2,3}$\thanks{E-mail: Ioannis.Apergis@warwick.ac.uk (IA)}
Daniel~Bayliss$^{\orcidlink{0000-0001-6023-1335}}$,$^{1,2}$
Paul~Chote,$^{1,3}$ 
James~McCormac$^{\orcidlink{0000-0003-1631-4170}}$,$^{1,2,3}$
Peter~J.~Wheatley$^{\orcidlink{0000-0003-1452-2240}}$,$^{1,2}$
\newauthor
Morgan A. Mitchell$^{\orcidlink{0009-0004-6130-7775}}$,$^{1,2}$
Jorge~Fernández~Fernández$^{\orcidlink{0000-0002-1416-2188}}$,$^{1,2}$
Sam~Gill$^{\orcidlink{0000-0002-4259-0155}}$,$^{1,2}$
Edward~M.~Bryant$^{\orcidlink{0000-0001-7904-4441}}$,$^{1,2}$
\newauthor
Toby~Rodel$^{\orcidlink{0009-0009-2175-72841}}$,$^{4}$
Leonidas~Asimakoulas$^{\orcidlink{0000-0003-3676-5606}}$,$^{5}$
David~R.~Anderson$^{\orcidlink{0000-0001-7416-7522}}$,$^{6}$
James~A.~Blake$^{\orcidlink{0000-0002-5903-2387}}$,$^{1,2,3}$
Sarah L. Casewell,$^{7}$
\newauthor
Fintan Eeles-Nolle$^{\orcidlink{0009-0009-6207-3217}}$,$^{1,2}$
Faith~Hawthorn$^{\orcidlink{0000-0002-8675-182X}}$,$^{1,2}$
James S. Jenkins$^{\orcidlink{0000-0003-2733-8725}}$,$^{8,9}$
Monika~Lendl$^{\orcidlink{0000-0001-9699-1459}}$,$^{10}$
Isobel~S.~Lockley$^{\orcidlink{0009-0003-0928-3588}}$,$^{1,2}$
\newauthor
Maximiliano~Moyano,$^{7}$
Sean~M.~O'Brien$^{\orcidlink{0000-0001-7367-1188}}$,$^{4}$
Suman Saha$^{\orcidlink{0000-0001-8018-0264}}$,$^{8,9}$
Alexis~M.~S.~Smith$^{\orcidlink{0000-0002-2386-4341}}$,$^{11}$
Philip~G.~Steen,$^{5}$
\newauthor
Jose~I.~Vines,$^{6}$
Richard~G.~West$^{\orcidlink{0000-0001-6604-5533}}$,$^{1,2}$
Tafadzwa~Zivave$^{\orcidlink{0009-0001-8055-995X}}$$^{1,2}$
\\
$^{1}$Department of Physics, University of Warwick, Gibbet Hill Road, Coventry CV4 7AL, UK\\
$^{2}$Centre for Exoplanets and Habitability, University of Warwick, Gibbet Hill Road, Coventry CV4 7AL, UK\\
$^{3}$Centre for Space Domain Awareness, University of Warwick, Gibbet Hill Road, Coventry CV4 7AL, UK\\
$^{4}$Astrophysics Research Centre, School of Mathematics and Physics, Queen’s University Belfast, Belfast, BT7 1NN, UK\\
$^{5}$Andor Technology Ltd, Springvale Business Park, 7 Millennium Way, Belfast BT12 7AL, UK\\
$^{6}$Instituto de Astronom\'ia, Universidad Cat\'olica del Norte, Angamos 0610, 1270709, Antofagasta, Chile\\
$^{7}$School of Physics and Astronomy, University of Leicester, University Road, Leicester LE1 7RH, UK\\
$^8$Instituto de Estudios Astrofísicos, Facultad de Ingeniería y Ciencias, Universidad Diego Portales, Av. Ejército Libertador 441, Santiago, Chile\\
$^9$Centro de Excelencia en Astrofísica y Tecnologías Afines (CATA), Camino El Observatorio 1515, Las Condes, Santiago, Chile\\
$^{10}$Observatoire de Gen{\`e}ve, Universit{\'e} de Gen{\`e}ve, Chemin Pegasi, 51, 1290 Versoix, Switzerland\\
$^{11}$Institute of Space Research, German Aerospace Center (DLR), Rutherfordstr. 2, 12489, Berlin, Germany\\
}

\date{Accepted XXX. Received YYY; in original form ZZZ}

\pubyear{2025}

\begin{document}
\label{firstpage}
\pagerange{\pageref{firstpage}--\pageref{lastpage}}
\maketitle

\begin{abstract}
Modern scientific CMOS cameras offer very fast readout speeds and low read noise. In this study, we evaluate the performance of the Andor Marana CMOS camera through on-sky testing carried out at the NGTS facility at the ESO Paranal Observatory in Chile. We mount the Marana camera to an NGTS telescope, and conduct photometric observations of bright stars. In particular, we target transit events around eight known bright exoplanet host stars. Simultaneous observations are carried out using an existing Andor iKon-L CCD camera on a neighbouring NGTS telescope. This allows for a direct comparison of the photometric precision between the CMOS and CCD cameras. We find that the Marana CMOS exhibits a similar level of photometric performance to the CCD camera, achieving 500\,ppm at a 30-minute timescale for a T $=10$\,mag star. Although the CCD has a slightly better quantum efficiency over the NGTS filter range (520-890\,nm), we find that the faster readout speed of the CMOS compared to the CCD means that the CMOS camera detects 20\,\% more photons per unit time for a solar-type star in our standard 10\,s exposure time operation mode. This results in the CMOS performing slightly better photometry in the photon-limited regime. We conclude that modern CMOS cameras, such as the Marana, are very well-suited for astronomical time-series photometry applications.
\end{abstract}




\section{Introduction}
\label{introduction}
Charge Coupled Device (CCD) cameras are an essential tool in modern astronomy, being at the heart of almost all astronomical instruments operating at optical wavelengths \citep{1986ARA&A..24..255M}. 

The major space-based exoplanet transit detection missions, such as Kepler \citep{2010Sci...327..977B} and the Transiting Exoplanet Survey Satellite \citep[\tess]{2015JATIS...1a4003R} have been equipped with CCD detectors. The upcoming PLATO mission \citep{plato} will also use CCD detectors. Similarly, ground-based transiting exoplanet surveys such as the Wide Angle Search for Planets \citep[WASP]{2006PASP..118.1407P} and its successor, the Next Generation Transit Survey \citep[\ngts]{2018MNRAS.475.4476W} have used CCD detectors. However rapid progress in Complementary Metal-Oxide Semiconductor (CMOS) technology, particularly over the past 10 years such as low read noise and extended dynamic range  mean that CMOS detectors may now be considered for future transiting exoplanet surveys \citep{2022arXiv220606693G}.


CMOS cameras provide substantial advantages for dynamic celestial observations requiring rapid data capture and low read noise, such as detecting exoplanet transits \citep{2018MNRAS.475.4476W}, tracking asteroids \citep{alarcon2023scientific}, monitoring meteoroids \citep{kajino2019study}, detecting trans-Neptunian objects \citep{2024SPIE13096E..2YW} and conducting space situational awareness observations to track and characterise satellites in near-Earth orbit \citep{zimmer2016affordable, blake2023exploring, cooke2023simulated, 2025arXiv250212324A, 2025AdSpR..76..764C}. 
Recent tests and observational campaigns have demonstrated the capability of CMOS sensors to produce high-quality photometry across a range of applications \citep{alarcon2023scientific,2024RAA....24e5009M}. CMOS cameras can achieve photometric precision comparable to that of CCDs, with accuracies reaching the milli-magnitude level \citep{2025arXiv250312449X}. On-sky and detector noise sources measurements have been found to be in agreement with the theoretical noise models \citep{2025arXiv250200101L}. In particular, modern back-illuminated CMOS detectors do not suffer the strong sub-pixel sensitivity variations associated with front-illuminated detectors (electrode structures, micro-lenses etc). \citep{10.1117/12.2561834}.

In this work, we set out to assess the performance of a CMOS camera in this context by deploying one at the NGTS facility, which is situated at a site with exceptional photometric qualities. We utilised a modern CMOS camera, the Marana 4.2BV-11, which has demonstrated great performance in laboratory tests \citep{2021RMxAC..53..190K, 2021RAA....21..268Q, Ioannis24} and is expected to be well-suited for high-precision time-series photometry. By directly comparing the CMOS camera's performance with that of Andor's iKon-L CCD camera that is already installed on the NGTS telescopes, we test if CMOS technology can deliver comparable bright star photometry to the well-tested CCD cameras. This study provides valuable insight into whether CMOS image sensors could potentially compete with CCD image sensors in exoplanet surveys. By evaluating if CMOS devices exhibit similar noise characteristics and performance, we aim to determine their suitability as trusted instruments for precise, stable photometric measurements.

Our first paper \citep{Ioannis24} evaluated the performance of the Marana CMOS camera under controlled laboratory conditions. We measured the camera’s read noise in dark environments and characterized both the temporal and spatial noise patterns across sensor rows and columns. We examined the behaviour of the dark current and glow at different operating temperatures. Under illuminated conditions, we quantified the photon response non-uniformity and assessed the camera’s linearity performance. Additionally, we measured the quantum efficiency of the sensor across multiple wavelengths and determined the transmittance of the front window mounted on the camera. Those results are summarized in \autoref{tab:specs_results}. Overall, the performance of the Marana camera demonstrates that it is well suited for high-precision time-series photometry.

In this second paper, we focus on on-sky testing of the Marana CMOS, and making a direct comparison to the iKon-L CCD cameras. To do this, we take observations simultaneously with the CMOS and CCD cameras, with identical hardware configurations for the mount, telescope, filter, and with the same sky conditions. An identical pipeline is used to perform data reduction and photometry routines for both cameras. 

This paper reports on on-sky tests of the Marana CMOS with the NGTS facility. It is organised as follows, in Section~\ref{sec:cameras_telescopes} we describe the NGTS facility and the CMOS and CCD cameras. In Section~\ref{sec:methods} we set out the methodology, including the installation of the CMOS camera at NGTS, the on-sky observations made with the CMOS and CCD cameras, and the image reduction and photometry pipeline. In Section~\ref{sec:noise} we discuss the various sources of photometric noise that are present in our CMOS and CCD data. In Section~\ref{sec:results} we present our photometric results for the CMOS and CCD cameras from the on-sky testing. In Section~\ref{sec:discussion} we discuss the results from the CMOS on-sky testing. Lastly, Section~\ref{sec:conclusion} outlines our conclusions on the suitability of the CMOS camera for bright star photometry. 

\section{Observatory and Instruments}
\label{sec:cameras_telescopes}
\subsection{The NGTS Facility}
\label{ngts}

The Next Generation Transit Survey \citep[NGTS]{2018MNRAS.475.4476W} is a ground-based observatory located at ESO’s Paranal Observatory in Chile. Inspired by the success of the WASP programme \citep{2006PASP..118.1407P}, NGTS was built to provide very high precision time series photometry for bright stars to detect and characterise transiting exoplanets.  

\begin{figure}
\includegraphics[width=\columnwidth]{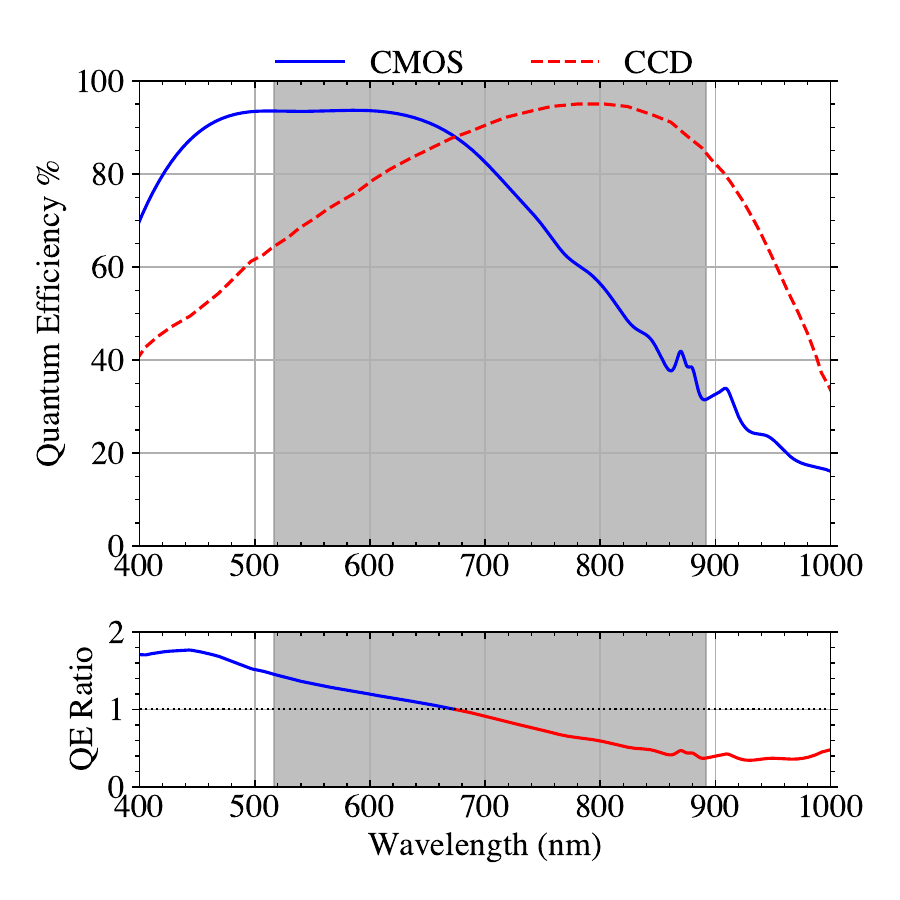}
\caption{\textbf{Top:} QEs  of the cameras as a function of wavelength. The filled grey area shows the NGTS custom filter. The blue line is the QE for the Marana CMOS camera and the red dashed line is the QE of the iKon-L CCD. \textbf{Bottom:} The ratio of the QE (CMOS/CCD).}
\label{fig:filters_ratio}
\end{figure}

NGTS is comprised of 12 Newtonian telescopes, each with $f$/2.8 and 20\,cm apertures, manufactured by AstroSysteme Austria\footnote{\url{https://www.astrosysteme.com/references/observatory-paranal-chile/}}. The telescopes are spaced 2\,m apart, in two rows of six, in a single enclosure. The optical design of the telescopes include a primary hyperbolic mirror along with a corrector lens. The telescopes are equipped with deep-depleted CCD cameras with high sensitivity in the red spectrum (see Section~\ref{sec:ikon}). These CCD cameras are mounted on a customised focuser with an NGTS-specific filter, designed for the 520-890\,nm range. The quantum efficiency of the cameras is shown in \autoref{fig:filters_ratio}. Further details about the instruments and the operation can be found in \citet{2018MNRAS.475.4476W}. 

The NGTS facility has successfully discovered 36 exoplanets and 8 eclipsing binaries\footnote{\url{https://ngtransits.org/pages/tabs/publications_page.html}}. Notable discoveries include NGTS-1b \citep{2018MNRAS.475.4467B}, a rare giant planet around a low mass star, NGTS-4b \citep{ 2019MNRAS.486.5094W}, the shallowest ever transit to be initially detected from a ground-based observatory, NGTS-10b \citep{2020MNRAS.493..126M}, a very short-period hot Jupiter, NGTS-14Ab \citep{2021A&A...646A.183S}, an exoplanet in the Neptunian desert and NGTS-30b \citep{2024A&A...686A.230B}, a very young long-period warm Jupiter. NGTS photometry has been shown to be scintillation limited \citep{2022MNRAS.509.6111O} for bright stars $\mathrm{T}<12 $ mag. The scintillation noise has been proven to be uncorrelated between the telescopes and therefore NGTS is capable of reaching the precision of TESS for stars of around $\mathrm{T}=9$\,mag \citep{2020MNRAS.494.5872B} when combining data from multiple telescopes.

Currently NGTS is focused primarily on discovering long period transiting exoplanets based on TESS mono-transits or duo-transits \citep{2020MNRAS.491.1548G, 2020ApJ...898L..11G, 2022A&A...666A..46U, 2024MNRAS.528.1841H}. Additionally, NGTS observes targets from a range of science areas, including bright exoplanet host stars \citep{2020Msngr.181...28B, 2022SPIE12191E..1AB}, targeted open cluster surveys \citep[e.g.][]{2020MNRAS.492.1008G}, long period eclipsing binaries \citep[e.g.][]{2020MNRAS.495.2713G, 2025MNRAS.537...35R}, transiting brown dwarfs \citep[e.g.][]{2024MNRAS.530..318H} and stellar flares \citep[e.g.][]{2023MNRAS.525.1588J}.

\subsection{The iKon-L 936 CCD Camera}
\label{sec:ikon}
The iKon-L 936 BR-DD is a CCD camera\footnote{\url{https://andor.oxinst.com/products/ikon-xl-and-ikon-large-ccd-series/ikon-l-936}} that was developed by Andor Technologies. It features a back-side deep depleted sensor with 2048 $\times$ 2048 pixels with pixel dimension 13.5\,$\unit{\um}$ each. The sensor is made by e2v and it is enhanced with fringe suppression and dual anti-reflection coatings.

The iKon-L CCD serves as the primary camera for NGTS, mounted on all NGTS telescopes. It provides high sensitivity in the red wavelengths, achieving QE $>90\,\%$ at 750\,nm (see \autoref{fig:filters_ratio}), which is particularly advantageous for observing K and M dwarf stars.  Equipped with a five-stage Peltier thermoelectric cooling system, the CCD maintains minimal dark current at operating temperatures as low as -100$^\circ$\,C. Additionally, it features vacuum sealing, protecting it from environmental influences and enabling long-term operation without the need for replacement or servicing.  

The camera underwent extensive laboratory testing \citep{2018MNRAS.475.4476W} (see \autoref{tab:specs_results}) before NGTS commissioning and successful deployment in 2015. NGTS operates the iKon-L CCD camera in the 3\,MHz readout mode and at -70\,$^\circ$\,C constant temperature. Typically NGTS uses 10\,s exposure time for imaging followed by 1.5\,s readout. However, additional delays are introduced due to autoguiding corrections and per-image statistical calculations. Furthermore, the mount requires a short settling period before the shutter can reopen. Altogether, these factors result in a total dead time between the images of approximately 3\,s.

\subsection{The Marana CMOS Camera}
\label{sec:marana}
The Marana 4.2B-11 Camera\footnote{\url{https://andor.oxinst.com/products/scmos-camera-series/marana-scmos}} is a CMOS device featuring a thinned, back-side illuminated sensor. The sensor has 2048 $\times$ 2048 pixels, with pixel dimension 11\,$\unit{\um}$. It offers low read noise, an extended dynamic range thanks to the high dynamic range (HDR) mode, and high sensitivity across the visible spectrum, with quantum efficiency peaking at 95$\,\%$ at 550 nm (see \autoref{fig:filters_ratio}). The sensor is housed within a vacuum chamber, which helps preserve sensor quality by effectively preventing condensation and dust accumulation. The five stage Peltier thermoelectric cooler enables operating temperatures down to -25\,$^\circ$\,C air-cooled. The camera also provides fast readout speeds and high frame rates, enabling high-cadence monitoring. The performance of the Marana CMOS is summarized in \autoref{tab:specs_results}.

For the on-sky testing at NGTS presented in this work, the CMOS camera was operated in High Dynamic Range (HDR) mode. This is the default mode for astronomical applications, offering low read noise and an extended dynamic range. The extended dynamic range and low read noise are achieved through the implementation of two pixel-level pre-amplifiers: a High Gain and a Low Gain. The signal from each pixel is processed simultaneously along both gain paths, with the High Gain channel optimised for low illumination levels and the Low Gain channel for high illumination levels. Each gain channel is digitised using a 12-bit analogue-to-digital converter \citep{ma2013low, tang2020high, lou2023over}. At the start of an exposure, once the High Gain channel becomes saturated, the system switches to the Low Gain channel (see further discussion in \autoref{sec:camera_efficiency}). The final output frame is reconstructed by combining the two gain channels into a single 16-bit image, producing a HDR output \citep[see][for details]{Ioannis24}. 

We operate the camera at a constant temperature of -25\,$^\circ$\,C. Based on the temperature information recorded in the image headers, no significant fluctuations were observed, and the detector temperature remained stable at the set point throughout the observations. We set the exposure time to be 10\,s in order to match that of the iKon-L CCD camera. 

The Marana camera applies internal pre-processing corrections intended to improve image quality. These include Anti-glow correction, which mitigates glow artefacts visible in dark exposures, and the Spurious Noise Filter, which replaces noisy pixels by averaging neighbouring pixels \citep[see][for details]{Ioannis24}. For the analysis presented here, the Spurious Noise Filter was disabled, while the Anti-glow correction was enabled.

\begin{table}
\caption{Specifications of the Marana CMOS camera \citep{Ioannis24} in comparison with the iKon-L 936 CCD \citep{2018MNRAS.475.4476W}}
\label{tab:specs_results}
\begin{tabular}{lcc}
\hline
\textbf{Camera}                                & \textbf{CMOS}      & \textbf{CCD}\\ \hline                        Model                                                       & Marana 4.2B-11  & iKon-L 936\\
Digital output                                       & HDR, 16-bit       & 16-bit\\
Readout modes                                        & 100 MHz           & 3MHz\\
Gain e$^-$/ADU                                       & 1.131             & 1.99  \\
Full Well Capacity, e$^-$pix$^{-1}$                  & 69026             & 82000\\
Non-Linearity, \%                                    & 0.122             & 1\\
Read Noise, e$^-$pix$^{-1}$                          & 1.571             & 12.9\\
Dark Current, \eps                                   & 1.617@-25$^{\circ}$C  & 0.005@-70$^{\circ}$C\\
PRNU*, \%                                            & 0.131             & 1.79\\
DSNU**, \,e$^-$                                      & 0.232             & 3.79\\  
Readout time, \,s                                    & 0.042             & 1.5   \\  
Detector size, \,mm                                  & 22.5$\times$22.5  & 27.6$\times$27.6\\  
Pixel size, \,$\unit{\um}$                           & 11                & 13.5 \\
NGTS Pixel scale, \,arcsec pix$^{-1}$                & 4.01              & 5.01\\ 
NGTS Field Of View, \,deg                            & 2.27$\times$2.27  & 2.84$\times$2.84\\
Shutter                                              & Shutterless       & 45\,mm Mechanical \\
Weight, Kg                                           & 3.0               & 4.6 \\ \hline
 \multicolumn{3}{l}{* Photo Response Non-Uniformity, ** Dark Signal Non-Uniformity}\\
\end{tabular}
\end{table}

\section{Methods}
\label{sec:methods}

\subsection{Installation of the Marana CMOS Camera at NGTS}
\label{sec:software_hardware_methods}
\begin{figure}
\includegraphics[width=\columnwidth]{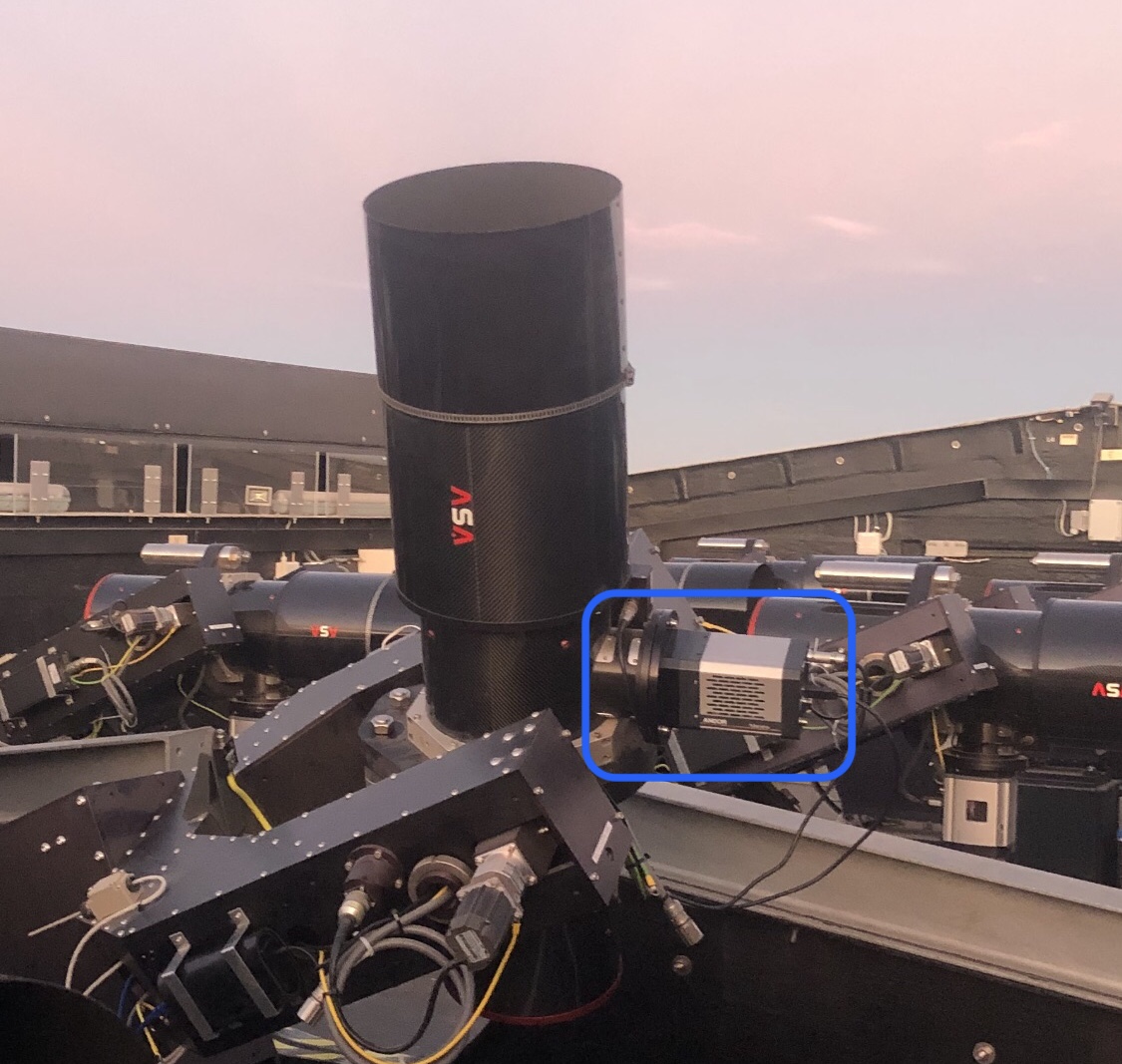}
\caption{The Marana sCMOS camera, highlighted within the blue rectangle, is mounted on one of the NGTS telescopes while capturing flat-field images during evening twilight at Cerro Paranal, Chile.}
\label{fig:marana}
\end{figure}

In December 2023 we installed a Marana CMOS Camera on one of the NGTS telescopes (telescope ID 06). For this installation we removed the existing iKon-L CCD from the telescope and mounted the Marana CMOS using a custom camera faceplate to achieve the correct telescope back focus distance. As the Marana CMOS is 1.6\,kg lighter than the iKon-L CCD camera, the telescope was rebalanced, which was achieved by removing some of its counterweights. \autoref{fig:marana} shows the Marana CMOS camera mounted on the NGTS telescope.

To operate the Marana CMOS camera on NGTS, we installed an Intel NUC 11 Extreme Kit Core control computer in a small cabinet next to the pier of the telescope. This control computer was equipped with an i9 CPU, 32\,GB RAM and 4\,TB SSD running the \texttt{Rocky~9} Linux operating system. The Marana CMOS camera was connected to the control computer via a USB 3.0 connection. This control computer allowed us to operate the Manara CMOS camera independently and without affecting the other eleven NGTS telescopes and cameras.  

The control computer operated both telescope and camera using the Robotic Observatory Control kit (\texttt{Rockit})\footnote{\url{https://github.com/rockit-astro}}, a software package designed for managing ground-based robotic observatories. \texttt{Rockit} is also used at the University of Warwick’s facilities at the Roque de los Muchachos Observatory \citep[e.g.][]{2025arXiv250212324A, mitchell_search_2025}. It follows a modular architecture in which background daemons run continuously to handle specific tasks such as telescope control, focusing, instrument management, and overall observatory coordination. These daemons seamlessly interact with each other through a remote procedure call interface, enabling the execution of functions and procedures on remote programs. To facilitate this communication, \texttt{Rockit} uses the python library \texttt{Pyro4}.


Similarly to NGTS, to maintain the stability of the stars in the images and minimise systematic errors and noise from tracking inaccuracies, we use the \texttt{DONUTS} algorithm \citep{2013PASP..125..548M}. This innovative python package functions as an autoguiding algorithm, maintaining pixel drift at an accurate sub-pixel level relative to the first image, which serves as the reference frame. This approach meets our high precision photometry requirements and ensures consistent software performance between the CMOS and CCD cameras.




\subsection{Observations}
\label{sec:observations}
Observations were conducted from December 2023 to September 2024, during which time the CMOS camera was mounted on one of the NGTS telescopes. We also utilised one of the eleven available CCD cameras on a neighbouring telescope (telescope ID 01) for simultaneous observations. To maintain consistency and ensure photometric stability, the same CCD camera was used for comparison with the CMOS camera throughout the entire campaign. We made observations of well-known transiting exoplanets around bright stars, including WASP \citep{2006PASP..118.1407P} and KELT \citep{2004AIPC..713..185P}  host stars, as well as confirmed TOI systems.  The targets are shown in \autoref{tab:targets}. For the remainder of the time, the CMOS camera and telescope observed the same fields as the CCD on telescope ID 01, which were primarily TESS monotransit and duotransit candidates monitored as part of the NGTS program to discover long period exoplanet systems from TESS \citep{2024MNRAS.528.1841H, 2024MNRAS.533..109G}.

The nightly schedule for the CCD on Telescope ID 01 was replicated and scheduled in the \texttt{Rockit} software to ensure the CMOS camera on Telescope ID 06 observed the exact same field centre as the CCD camera. This setup guaranteed that both cameras monitored the same area of sky simultaneously, although we note the CMOS camera has a smaller field-of-view (see \autoref{tab:specs_results}). This strategy maintained consistency in observation duration and coverage between the CMOS and the CCD image sensors which we use for comparisons in this work. The CCD camera operated with an exposure time of 10 seconds and a cadence of 13 seconds. For consistency, the CMOS camera was configured with a 10 second exposure time and a total cadence of 10.042 seconds. NGTS uses a custom filter that covers a wavelength range of 520-890\,nm (see Section~\ref{sec:cameras_telescopes} and \autoref{fig:filters_ratio}).  We used this same filter for the observations made using the CMOS camera, although we note the filter is not optimized for the bluer CMOS sensor's quantum efficiency (see \autoref{fig:filters_ratio}). The images from the Marana CMOS camera were saved as FITS files onto the CMOS control computer. Instrument parameters were saved into the FITS file headers, including the JD timestamps, telescope coordinates, and camera settings.  

In this paper we analyse data from nights where known transiting exoplanet systems were observed. We observed eight transiting exoplanets in total, with three of them observed twice. We also analyse two nights of observations taken on a dark night with no moon (2024 July 5) and a bright night with full moon (2024 June 22), both centred on the star TIC-188620407 (R.A.= 23$^{\mathrm{h}}$20$^{\mathrm{m}}$12\fs37, Dec= -13$^{\mathrm{\circ}}$5\arcmin56\farcs83, T $=11.57$\,mag).  

\begin{table*}
\centering
\caption{Simultaneous transit observations of bright stars with CMOS and CCD cameras.}
\begin{tabular}{ccccccc}
\hline
\hline
Target   & TIC         & TESS            & Teff             & Observation        & Observation Duration & Transit fraction \\ 
   &  ID        &  Magnitude *          &  (K) *           &   Date      & (hours) & \\ 
\hline
TOI-905 \citep{2020AJ....160..229D}  & TIC 261867566 & 10.57  & 5380 $\pm$ 130   & 2024-Jul-13 & 5.7    & Full     \\
WASP-95 \citep{2014MNRAS.440.1982H}  & TIC 144065872 & 9.49   & 5760 $\pm$ 130   & 2024 Jul 14 \& Aug 8 & 7.4 \& 8.8   & Full    \\
TOI-2109 \citep{2021AJ....162..256W} & TIC 392476080 & 9.78   & 6650 $\pm$ 130   & 2024 Jul 17 & 5.5    & Full     \\
KELT-10 \citep{2016MNRAS.459.4281K}  & TIC 269217040 & 10.27  & 5890 $\pm$ 130   & 2024 Aug 01 \& 26 & 8.3 \& 6.5    & Ingress     \\
WASP-4  \citep{2008ApJ...675L.113W}  & TIC 402026209 & 11.82  & 5500 $\pm$ 140   & 2024 Aug 02 & 7.4    & Full     \\
WASP-97 \citep{2014MNRAS.440.1982H}  & TIC 230982885 & 9.96   & 5700 $\pm$ 130   & 2024 Aug 04 \& 6 & 5.6 \& 5.7   & Egress \& Full     \\
TOI-4463 \citep{2023ApJS..265....1Y} & TIC 8599009   & 10.46  & 5540 $\pm$ 140    & 2024 Aug 27 & 4.2   & Full     \\
WASP-30 \citep{2011ApJ...726L..19A}  & TIC 9725627   & 10.97  & 6360 $\pm$ 130    & 2024 Aug 28 & 7.9   & Full     \\ \hline

 \multicolumn{3}{l}{* source: TIC v8.2 \citep{2018AJ....156..102S}}\\
\label{tab:targets}
\end{tabular}
\end{table*}

\subsection{Image reduction and Photometry pipeline}
\label{sec:photometry_pipeline_methods}
For each observation,  we execute a custom pipeline daily to extract photometry for each observed field. We solve each image astrometrically using the World Coordinate System (WCS), accounting for the field centre and pixel scale, with the help of \textit{astrometry.net} \citep{2010AJ....139.1782L}. This software solves each image using a simple polynomial model. However, we identify weaknesses in the solution near the image corners, caused by distortions from the telescope’s optics. To address this, we incorporate the TESS Input Catalog version 8 \citep[TIC8]{2018AJ....156..102S} to identify sources in the field, particularly those brighter than $\mathrm{T} < 16$\,mag. Using the initial WCS solution, we iteratively refine the polynomial fit until we achieve sub-pixel accuracy \citep[see][for details]{2025AdSpR..76..764C} . We exclude from the dataset any images for which the WCS solution fails; these are typically outliers.

For the selections of stars for photometry, we used TIC8 \citep{2018AJ....156..102S} with an upper cut of $\mathrm{T} < 14$\,mag. We apply criteria for blended stars, excluding stars that have a source identified within a 6 pixel radius (24 arcseconds) around the star and a maximum magnitude difference of 2.5 magnitudes for a star to be considered as blended. Additionally, we exclude stars near the edges of the sensor. The same extracted catalog for the CMOS image sensor is used for the CCD image sensor, since the FOV is larger due to the larger pixel scale, ensuring that both fields will contain the same stars. 

We acquired dark and bias images when we were on-site, using the telescope cover to prevent external light from reaching the sensor. The properties and the long-term stability of the CMOS camera's bias and dark frames are discussed extensively by \citet{Ioannis24}. We record evening and morning flat-field images before and after the science observations. We averaged the dark and bias frames to remove the electronic and thermal signal from each camera, and we create an averaged and normalized flat-field image for flat correction.

For the CCD camera, a flat-field correction was not applied, primarily because the autoguiding system keeps stellar positions stable to within sub-pixel deviations ($<$0.1\,pixel RMS in X and Y positions over a typical observing night), ensuring that the same pixels are used consistently throughout the night \citep{2013PASP..125..548M}. The CMOS camera also exhibits sub-pixel positional stability. We performed photometry both with and without applying flat-field corrections for the CMOS data and found no noticeable difference. For a typical sample of stars, we find that omitting flat-field correction results in an RMS scatter improvement of approximately 0.29\,\%. Therefore, we adopted the same approach as for the CCD camera and did not apply a flat-field correction.

As part of the pipeline, we also verify that the autoguiding system is performing well by measuring the offsets between images using the \texttt{DONUTS} software \citep{2013PASP..125..548M} to ensure that stars remain on the same pixels. We set an upper limit for the x-y positional shifts in the images; any frames exceeding this threshold—such as those affected by satellite trails or guiding lasers from the Very Large Telescope—are excluded from the data. In most cases, the autoguiding performed well within the specified pixel shift threshold.

\begin{figure}
\includegraphics[width=\columnwidth]{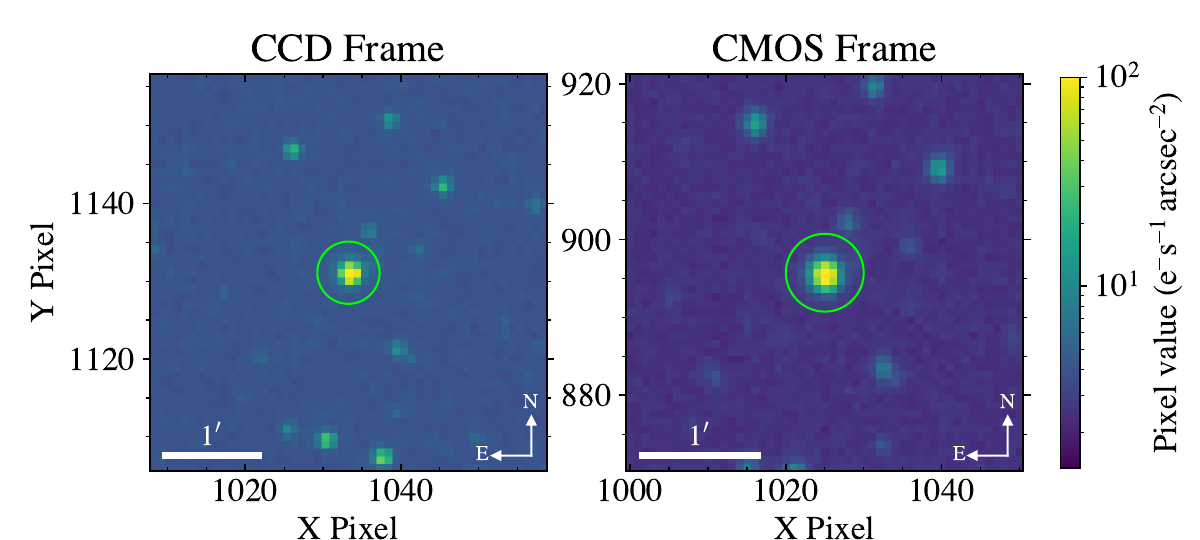}
\caption{Target pixel frame (50 $\times$ 50 pixels) for CCD on the left and CMOS on the right. The selected photometric apertures for each camera are shown with green circles for target star KELT-10 (TIC-269217040, T $=10.27$\,mag) observed on 2024 August 1.}
\label{fig:cmos_ccd_frames}
\end{figure}

To perform aperture photometry we extract the position for each star from the catalog. We use the \texttt{SEP} python package \citep{1996A&AS..117..393B, 2016JOSS....1...58B} with circular apertures of specific radii defined for each camera as set out in Section~\ref{sec:noise}. We define a sky background map from the \texttt{SEP} option to measure the sky background signal. An example of the target pixel frames for each camera is shown in \autoref{fig:cmos_ccd_frames}.
    
We perform relative photometry to correct the light curves for atmospheric extinction, color-dependent extinction, and any other systematics common to stars within the image. For each star, we collect its \textit{Gaia} DR3 $\mathrm{G_{BP}}-\mathrm{G_{RP}}$, TESS magnitude, and TIC ID. We define magnitude and color ranges for selecting comparison stars, with a bright magnitude limit of $T = 9.4$\,mag, which usually indicates saturation in a 10\,s image.


We establish criteria for the comparison stars by finding the best RMS value for each star within a particular magnitude range. Noisy stars are then excluded (see Section~\ref{sec:lc_comparison_results}). The remaining comparison stars are combined to produce a master reference star that corrects for the systematics affecting the target star.

\section{Photometric Noise}
\label{sec:noise}
There are several sources of photometric noise that are well known and can be calculated for both the CMOS and CCD cameras. We construct the photometric noise model using the read noise and dark current of each camera (see \autoref{tab:specs_results}), the sky background brightness and the target shot noise, and the expected scintillation noise at Paranal Observatory \citep{2022MNRAS.509.6111O}.  In order to be able to compare the CMOS and CCD cameras, we express the photometric noise in terms of electrons using the conversion factor (gain) from analog digital units to electrons (see \autoref{tab:specs_results}). Since the pixel size is different between the CMOS (11\,$\unit{\um}$) and the CCD (13.5\,$\unit{\um}$) image sensors, we calculate the photometric noise for a photometric aperture of a fixed physical radius of almost exactly 55\,$\unit{\um}$, which equates to a radius of 5 pixels for the CMOS image sensor and 4 pixels for the CCD image sensor. The typical full-width half-maximum (FWHM) of the NGTS point-spread-function is 19\,$\unit{\um}$, therefore a photometric aperture of 55\,$\unit{\um}$ was found to be optimal for our stars with magnitudes in the range $9<\mathrm{T}<11$ \citep{2000hccd.book.....H}.
 

\subsection{Detector Noise}
The two major sources of detector noise for digital imaging are read noise and dark current. Read noise is introduced during the process of the charge to voltage conversion and the conversion from analog to digital values. Read noise includes all the noise sources which are independent of the exposure and the incoming signal \citep{2000hccd.book.....H}. Traditionally, CMOS image sensors have exhibited higher read noise compared to CCDs, primarily due to the complex circuitry, the large number of amplifiers and their lower quality in earlier designs. However, advancements in CMOS image sensors such as the Marana camera have enabled the achievement of low read noise levels. The median read noise of the Marana camera was found to be 1.57\,e$^-$ \citep{Ioannis24}. Furthermore, while CCD image sensors struggle to maintain low read noise at high readout speeds, CMOS active pixel sensors are superior in this regard due to their unique amplifier-per-pixel design. For more details related to the read noise of the Marana CMOS camera see \citet{Ioannis24}.




Dark current is thermal noise produced by the generation of heat within the sensor and it is dependent on the selected exposure time and temperature of the detector \citep[e.g.][]{PhysRev.87.835, Aberle1992ImpactOI}. Unlike the passive pixel design of CCDs, CMOS image sensors employ active pixels to amplify the signal on a per-pixel basis \citep{stefanov2022cmos}. However, this additional circuitry can emit infrared photons from either the sensor periphery or the pixel amplifier transistor \citep{2007SPIE.6690E..03J, wang20154m}. This glow effect is distinguishable from the dark current at low temperatures, since it exhibits a weaker temperature dependence than the dark current \citep[see Figure 14 in][for details]{Ioannis24}. Consequently, as the detector temperature decreases, the relative contribution of the glow component becomes more significant. The dark current for the Marana camera was measured to be 1.617\,\eps at constant temperature of -25\,$^\circ$\,C. For further details on dark current and glow in the Marana CMOS camera, refer to \citet{Ioannis24}.




\subsection{Sky Background noise}
Sky background is a major source of photometric noise in astronomy, especially when monitoring faint objects. Sky background signal consists of all the light entering the photometric aperture other than light from the target object. This may be from scattered light, background astrophysical sources, zodiacal light, and atmospheric effects such as moonlight or air-glow from night-sky emission lines \citep{1993PASP..105..940M, wust2022hydroxyl}. The mean sky background signal is subtracted from the photometric aperture during the photometric extraction (see Section~\ref{sec:photometry_pipeline_methods}). However, the sky background noise will contribute to the Poisson photon noise. The CMOS and CCD cameras show differing amounts of sky background noise largely due to the differences in the quantum efficiency of the cameras over the wavelength range of the NGTS filter (see \autoref{fig:filters_ratio}).




\subsection{Source Noise}
\label{sec:source_noise}
For ground-based photometric observations, the noise contribution due to light from the target star has two components - Poisson shot noise and scintillation noise. Poisson shot noise from the target star is simply given by the square root of the measured brightness of the star in electrons.



Scintillation noise arises from the fact that incident light from the target star will fluctuate in intensity as it moves through turbulent regions in the Earth's atmosphere \citep{2015MNRAS.452.1707O}.  This effect causes the star to "twinkle". Scintillation is particularly noticeable for small aperture telescopes observing bright stars \citep{2022MNRAS.509.6111O}, where it becomes a dominant source of noise. We calculate the expected scintillation noise for NGTS using a modified version of Young’s formula \citep{1967AJ.....72..747Y, 2015MNRAS.452.1707O}:

\begin{equation}
    \mathrm{\sigma_{sc}^2} = 10^{-5} \mathrm{C_{Y}^2} D^{-4/3} t_{\mathrm{exp}}^{-1} \mathrm{sec}^3(z) \mathrm{exp}(-2h_{\mathrm{o} }/H),
\end{equation}

where $D$ is the diameter of the telescope (0.2\,m for NGTS), $\mathrm{t_{exp}}$ is the exposure time (10\,s for our observing), sec($z$) denotes the airmass, h$_\text{obs}$ is elevation above sea level (2433\,m for NGTS), and $H$ is the atmospheric scale height \citep[we adopt 8000\,m from][]{2015MNRAS.452.1707O}.  C$_\text{Y}$ is an empirical coefficient that quantifies the amplitude of scintillation on a given night. We adopt the median value for Paranal Observatory  of C$_\text{Y}=1.54\,\mathrm{m^{2/3}s^{1/2}}$ as reported by \cite{2022MNRAS.509.6111O}.

Unlike the Poisson shot noise from the star, the fractional scintillation noise is \textit{independent} of the flux from the target star \citep{2022MNRAS.509.6111O}. Thus for bright stars, scintillation noise will form a hard limit to the fractional photometric noise.



\subsection{Total Noise}
\label{sec:total_noise}
The total photometric noise is the quadrature sum of the individual noise sources for a given photometric aperture, as the noise will be statistically independent \citep{1989PASP..101..616H}: 

\begin{equation}
\label{eq.6}
    \mathrm{\sigma_{total}} = \sqrt{\mathrm{\sigma_{flux}}^2 + \mathrm{\sigma_{sky}}^2 + \mathrm{\sigma_{DC}}^2 + \mathrm{\sigma_{RN}}^2 + f^2\mathrm{\sigma_{sc}}^2},
\end{equation}

where $\mathrm{\sigma_{flux}}$ is the photon shot noise, $\mathrm{\sigma_{sky}}$ is the background signal noise, $\mathrm{\sigma_{DC}}$ is the dark current noise, $\mathrm{\sigma_{RN}}$ is the read noise, $f$ is the flux from the target star and $\mathrm{\sigma_{sc}}$ is the scintillation noise. In order to calculate the photometric noise expected for a star of a given magnitude, we must convert from flux in electrons ($f$) to magnitude ($m$) using the following equation:





\begin{equation}
\label{eq.7}
    \mathrm{m} = \mathrm{z_p} + 2.5\log(\mathrm{G}) - 2.5\log(\frac{f}{\mathrm{t_{exp}}})
\end{equation}

where $z_p$ is the photometric zeropoint, $\mathrm{G}$ is the conversion factor from electrons to digital units G (e$^-$ADU$^{-1}$) as listed in \autoref{tab:specs_results}, and $\mathrm{t_{exp}}$ the exposure time of the image. The zeropoint for the NGTS filter is determined by measuring the source flux of each star (in electrons) and comparing it to the known TESS magnitude of the star. Using the noise calculations and the conversion from flux to magnitude, we plot both the individual noise components and the total noise model in \autoref{fig:noise_dark_bright}.  



\subsection{Systematic Noise}
\label{sec:noises_timescale}
All of the noise sources set out in Section~\ref{sec:noise} produce uncorrelated (white) noise. However it is well known that usually there is some level of systematic noise in the photometry that will give rise to correlated (red) noise \citep{2006MNRAS.373..231P}.  

To quantify this, we test both cameras for correlated noise as a function of time. We bin the data at different timescales to examine the noise characteristics. \cite{2017PASP..129b5002M} demonstrated that NGTS photometry using the iKon-M CCDs showed very little red noise, as the majority of its contribution was mitigated by correcting for atmospheric extinction based on the airmass as well as the colour-dependent atmospheric extinction using \texttt{SysRem} python software \citep{2005MNRAS.356.1466T}.


We calculate the RMS of the lightcurves from unbinned cadence all the way up to 1800\,s binning cadence, and compare the values to the square-root N value expected for purely white noise (see Section~\ref{sec:timescale_results}).







\section{Results}
\label{sec:results}

\subsection{Noise model comparison}
\label{sec:noises_results}
To assess the on-sky performance of the cameras, we compared measured noise characteristics to the theoretical noise model, which includes read noise, dark current, sky background noise, target shot noise, and scintillation noise (see Section~\ref{sec:total_noise}).

Read noise and dark current are calculated using each camera's performance specifications (see Section~\ref{tab:specs_results}). To model the sky background signal, we calculated the median value of all pixels for all bias and dark subtracted images taken on a particular observing night. This provides a good estimate of the sky background for nights when the sky background is not significantly varying. For scintillation noise, we used the average airmass of the total observation and included locational parameters specific to the NGTS observatory (see Section~\ref{sec:source_noise}). Finally, we created a synthetic flux array to calculate the shot noise, normalizing all noise parameters with respect to that flux. We measured the zeropoint from the fluxes extracted from photometry and their corresponding TESS magnitudes. The zeropoint was used as in \autoref{eq.7} in order to convert to synthetic magnitudes. The total noise is the summation in quadrature of all the noise components as shown in \autoref{eq.6}. 

For our noise model comparison, we acquired a set of data of the same field (R.A.= 23$^{\mathrm{h}}$20$^{\mathrm{m}}$12\fs37, Dec= -13$^{\mathrm{\circ}}$5\arcmin56\farcs83) for a total of 5.5 hours on the night of 2024 June 22 (full moon) and on the night of 2024 July 5 (no moon). Moon conditions were selected in order to assess the performance of each camera with maximum and minimum sky background contribution. We employed relative photometry for stars as described in Section~\ref{sec:photometry_pipeline_methods} on a TESS magnitude range of $8<\mathrm{T}<14$, with the same size physical photometric aperture for each camera (see the introduction of \autoref{sec:noise}). This photometric aperture radius was found to be $\approx$2.9 $\times$ FWHM (1 $\times$ FWHM = 19.3\,$\unit{\um}$), which equates to a radius of 4 pixels for the CCD and 5 pixels for the CMOS \citep{2000hccd.book.....H}.

\begin{figure*}
\includegraphics[width=\textwidth]{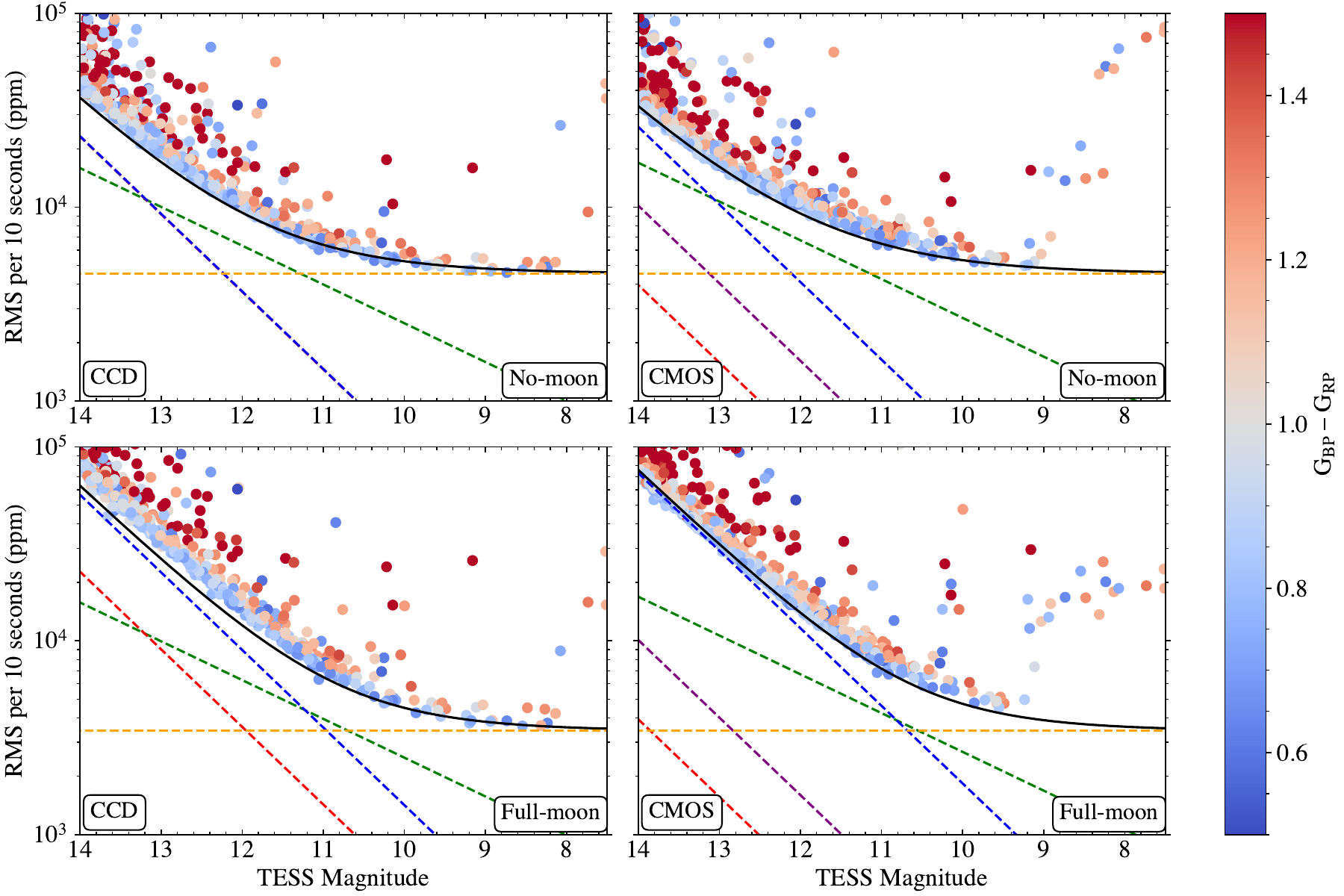}
\caption{Photometric precision as a function of stellar magnitude for 10 seconds exposure predicted by the theoretical model and from observations during the new moon on 5th July 2024 (upper panels) and full moon on 22nd June 2024 (lower panels). The left panels represent data from the CCD, while the right panels correspond to the CMOS. The dark line indicates the total modelled noise, including contributions from scintillation (yellow), photon shot noise (green), dark current noise (purple), sky background noise (blue), and read noise (red). The data points are coloured according to the \textit{Gaia} DR3 $\mathrm{G_{BP}}-\mathrm{G_{RP}}$ colour and represent the measured noise in the detrended light curves from relative photometry of 746 stars with $8<\mathrm{T}<14$ in a field centered on TIC-188620407.}
\label{fig:noise_dark_bright}
\end{figure*}

For each star we calculated the fractional RMS from the relative normalized lightcurves. The results for the datasets are shown in the upper panels (no moon) and lower panels (full moon) of \autoref{fig:noise_dark_bright}. We colour code the RMS of each star with the \textit{Gaia} DR3 $\mathrm{G_{BP}}-\mathrm{G_{RP}}$ colour to check RMS for a colour dependence.  

The RMS distribution of the stars is in good agreement with the calculated total noise model for both cameras and both datasets. In all cases we see that that reddest stars have a slightly higher RMS than bluer stars. This is likely to be due to the fact that we are plotting the TESS magnitudes for these stars (600-1000\,nm), but our photometry is in the NGTS band (520-890\,nm). The difference between these bandpasses means that redder stars appear brighter in the TESS filter compared with the NGTS filter, which does not include wavelengths beyond 890\,nm. Consequently, in \autoref{fig:noise_dark_bright}, the reddest stars would appear brighter using their TESS magnitudes, and fainter for our NGTS photometry. Thus, these redder stars have slightly higher RMS. Our relative photometry and detrending methods are also optimised for the average color star in the field, which will be much bluer than these very red stars. This could also result in systematics in the lightcurves of the redder stars which would produce higher RMS.



\begin{figure*}
\includegraphics[width=\textwidth]{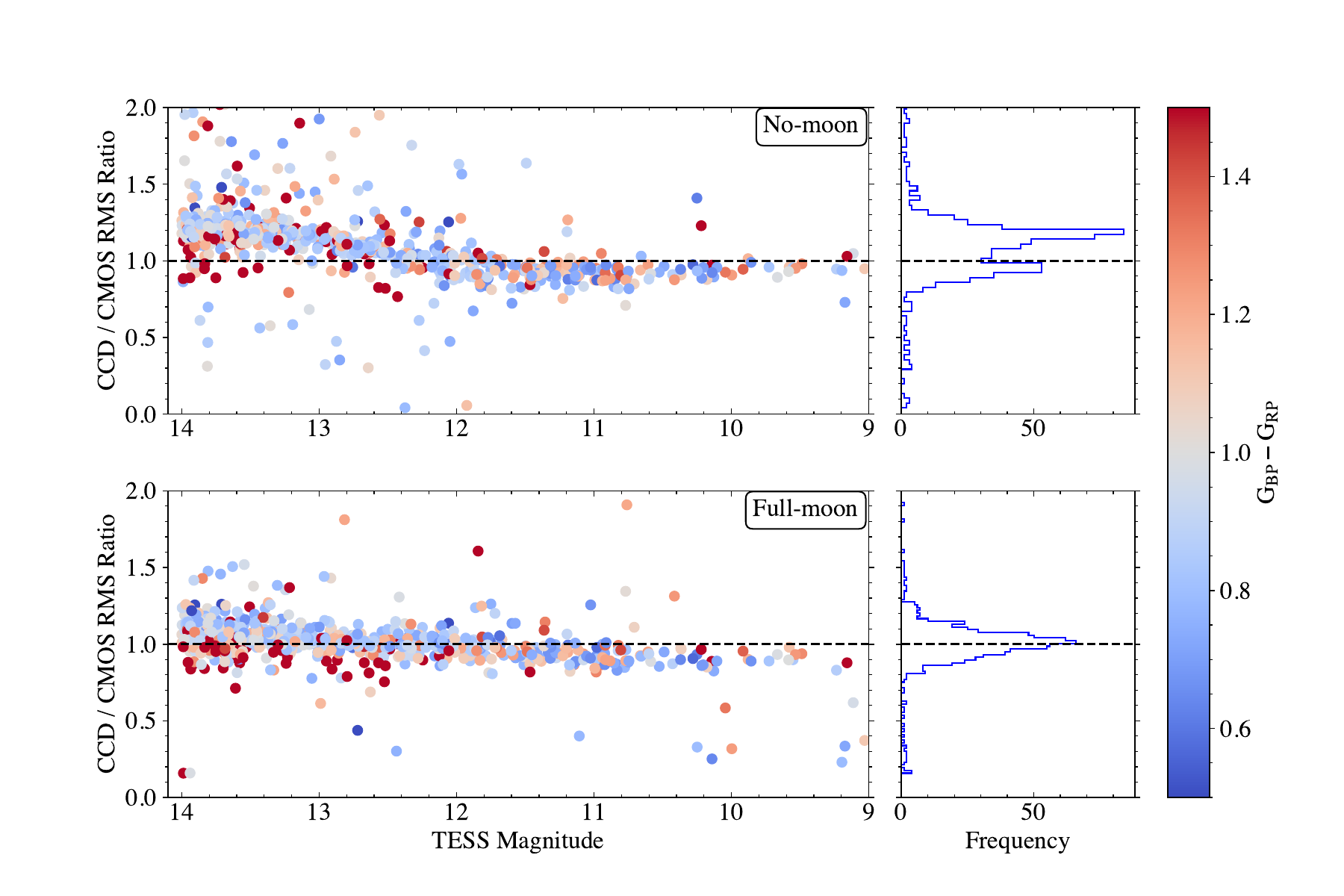}
\caption{Ratio for CCD and CMOS camera noise contribution for different magnitude stars from \autoref{fig:noise_dark_bright}. The data points are coloured according to the catalogued \textit{Gaia} DR3 $\mathrm{G_{BP}}-\mathrm{G_{RP}}$ colour. The black dashed line represents the ratio 1 which is the case where the noise is the same for both cameras. \textbf{Top:} The noise ratio for a dark night (no-moon). \textbf{Bottom:} The noise ratio for a bright night (full-moon). The panels on the right show the histogram distributions.}
\label{fig:rms_ratio}
\end{figure*}

To directly compare the RMS for given stars as simultaneously observed in each camera, we plot the ratio of CCD RMS to CMOS RMS for each star, as shown in \autoref{fig:rms_ratio}. When this RMS ratio exceeds unity, the RMS of the star is lower in the CMOS camera than in the CCD camera - i.e. the CMOS camera shows better precision in the lightcurve. Again we colour-code each star using its \textit{Gaia} DR3 $\mathrm{G_{BP}}-\mathrm{G_{RP}}$. 

\subsection{Timescale of noise}
\label{sec:timescale_results}
\begin{figure*}
\includegraphics[width=\textwidth]{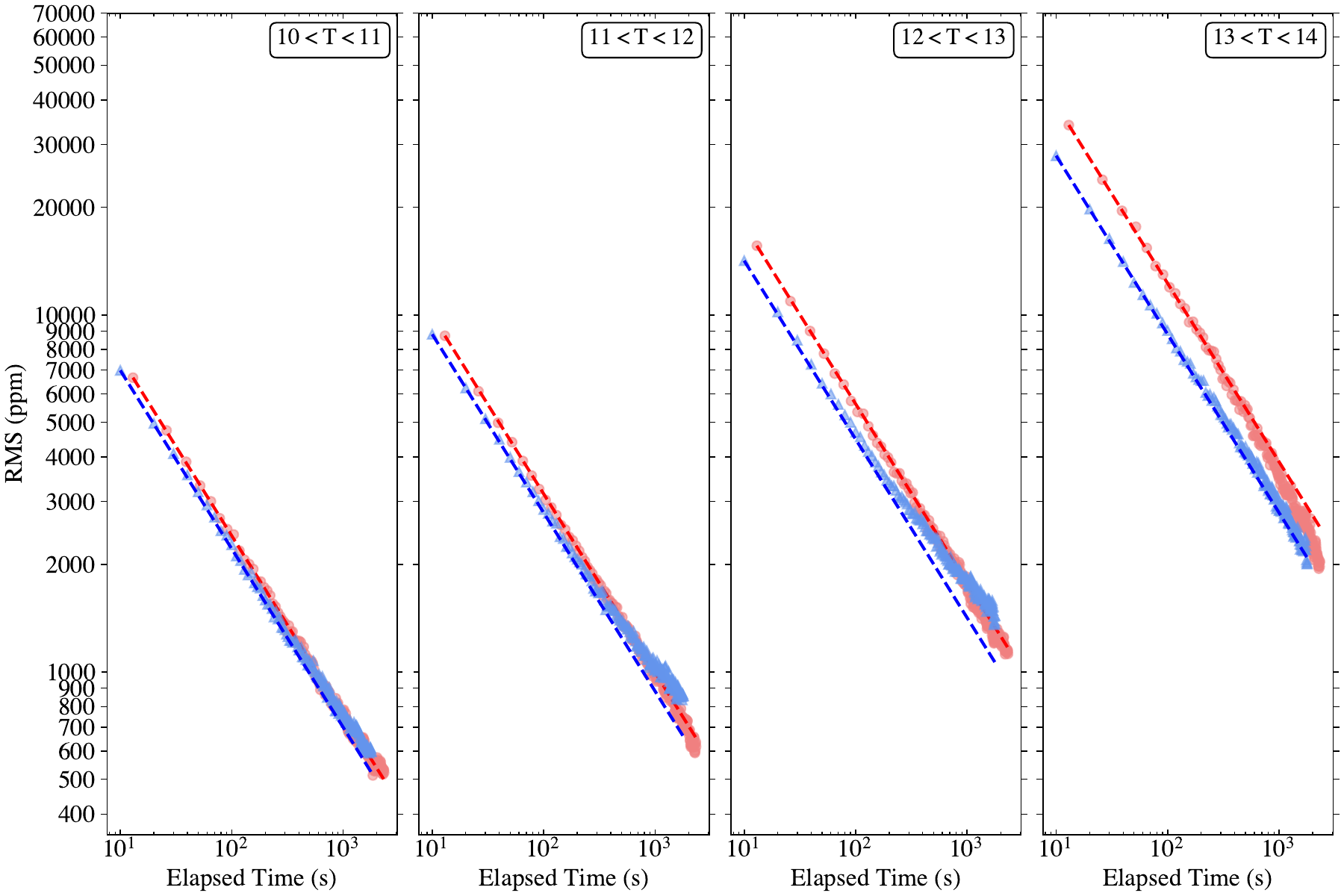}
\caption{Noise levels for CCD (light red circles) and CMOS (light blue triangles) at varying timescales during observations conducted on the no-moon night of 2024 July 5. The data is binned as a function of exposure and readout times. The blue and red dashed lines indicate the white noise model, proportional to $1 / \sqrt{n}$ for the CMOS and CCD respectively . The RMS noise was measured from stars that are non-variable and non-saturated. For the CMOS camera, 45 stars were randomly selected and averaged in each magnitude regime, and the same stars were used for the CCD camera.}
\label{fig:rms_timescale}
\end{figure*}

In order to investigate the presence of red noise in our data from each camera, we calculate the RMS for each lightcurve over a range of binning timescales for the observations taken with no moon (2024 July 5).  
We trim the beginning and end of each night’s data to include only observations taken at airmass < 1.7. This helps minimize excess noise associated with high airmass \citep{2020MNRAS.494.5872B}, while retaining enough data for effective binning. The total usable data amounted to approximately 4.8 hours. Stars were grouped in 1 TESS magnitude intervals, and their RMS was averaged and plotted for each camera, as illustrated in \autoref{fig:rms_timescale}. The maximum binning duration applied was 30 minutes. For each magnitude range, 45 stars were randomly selected, and the same set of stars was used for both cameras to ensure consistency. Each bin accounts for both the exposure, read and dead times. The exposure time was 10\,s for both cameras, while the readout times differed. For the CMOS camera the readout time is 42\,ms and we consider this negligible in our calculations. The CCD has a total of 3\,s read and dead time (see Section~\ref{sec:ikon}).

For the first two magnitude ranges, $10<\mathrm{T}<11$ and $11<\mathrm{T}<12$, the RMS values for both cameras are comparable, with the CCD camera exhibiting slightly lower noise, as also illustrated in \autoref{fig:rms_ratio}. In both cases, the RMS decreases similarly with binning time, following the expected trend from the white noise model. In the fainter magnitude bins, $12<\mathrm{T}<13$ and $13<\mathrm{T}<14$, the initial RMS for the CMOS camera is lower than that of the CCD camera, consistent with the relative performance indicated by the RMS ratio in \autoref{fig:rms_ratio}.


\subsection{Transit lightcurves}
\label{sec:lc_comparison_results}
A total of 11 successful nights of simultaneous observations of 8 well-known exoplanets transiting bright stars were conducted using the CMOS and CCD cameras, as summarised in \autoref{tab:targets}. Relative photometry, as outlined in Section~\ref{sec:photometry_pipeline_methods}, was employed to detrend the light curves from systematic effects. For each target star, an exclusion criterion was applied to remove comparison stars exhibiting high noise levels or significant systematics\footnote{\url{https://github.com/EMBryant/bsproc}}. 


We applied additional detrending to all 11 of our target lightcurves using \texttt{allesfitter}, a \texttt{Python} package \citep{allesfitter-code, allesfitter-paper}. To sample the parameters, we employ detrending using Gaussian Process on a covariance function, the Matern 3/2 kernel. The parameters are the characteristic amplitude $ln(\sigma)$ and length $ln(\rho)$ of scale.

To model the transits, \texttt{allesfitter} package integrates several specialised libraries such as \texttt{ellc} \citep{maxted2016ellc} for lightcurve modelling, \texttt{dynesty} \citep{speagle2020dynesty} for nested sampling \citep{Skilling2004nestedsampling, Skilling2006nestedsampling} and \texttt{celerite} for Gaussian processes \citep{foreman-mackey2017celerite}.


The 8 transiting exoplanet systems that we observed have already been extensively observed and characterised with high-precision data from both ground- and space-based instruments. Therefore we chose to fit a photometric model using nested sampling of the initial parameters with their uncertainties adopted from the most up-to-date existing literature as listed on the NASA exoplanet archive for each system \citep{2013PASP..125..989A}. We used normal priors for orbital period $P$, the radius ratio ($R_\text{B} / R_\text{A}$), semi-major axis ($(R_\text{A} + R_\text{B}) / a$), the epoch ($T_{0,\text{B}}$), orbital inclination ($\cos{i_B}$) and the quadratic limb darkening coefficients calculated based on the effective temperature, metallicity and the logarithm surface gravity ($log(g)$ of the host star \citep{2010A&A...510A..21S}). The 8 selected exoplanet system were all hot Jupiters with circular orbits reported in the literature. Therefore we set the eccentricity and argument of periastron of the system (implemented as the terms $\sqrt{e} \cos{\omega}$ and $\sqrt{e} \sin{\omega}$) as zero. All prior parameters were set to be the same for both the CCD and the CMOS cameras' extracted lightcurves. We did not include a dilution factor, as all the target stars are well isolated from nearby bright objects.

The results for the CCD and CMOS cameras' extracted lightcurves are plotted in \autoref{fig:WASP-4} for WASP-4 and in Appendix A as \autoref{fig:toi-905} to \autoref{fig:wasp-30} for all other targets. The upper panels in \autoref{fig:WASP-4} display the normalised lightcurves as extracted from the relative photometry and the detrending fit, the middle panels show the transit fit and the lower panels the residuals from the model. For plotting purposes we bin the detrended and model fitted lightcurves to 5 minutes bins (for both cameras) as shown in \autoref{fig:WASP-4}.

\begin{figure*}
\includegraphics[width=\textwidth]{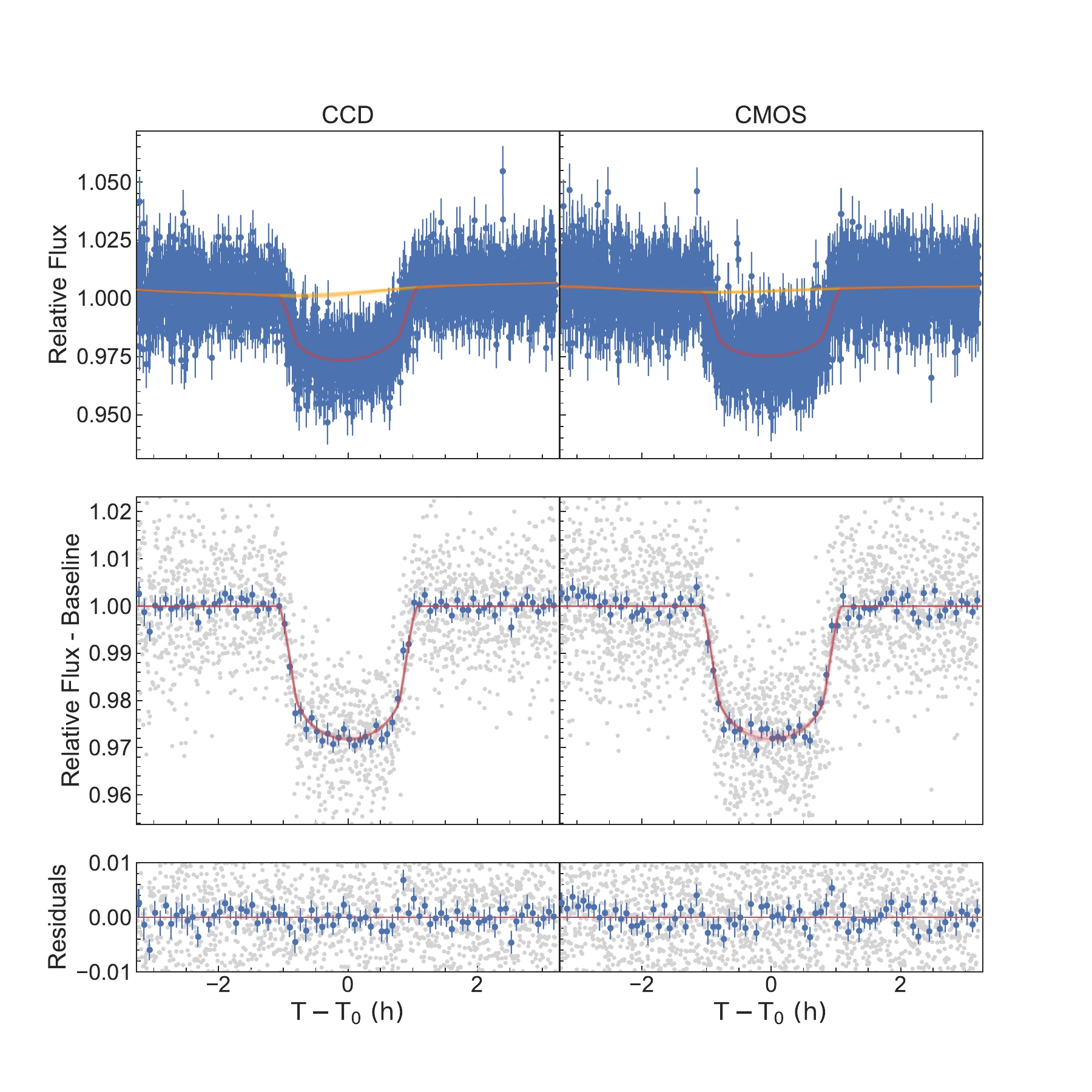}
\caption{Normalized lightcurves for the transiting system of WASP-4 (TIC 402026209) from NGTS on 2024 August 2 for CCD (\textbf{left}) and CMOS \textbf{right}) cameras.\textbf{Top}: Undetrended light curves with the GP detrending function shown as yellow solid line and transit model shown with red solid line. \textbf{Middle}: Detrended lightcurve with data binned to 5 minutes (solid blue circles) for clarity. Again the best fit model is shown in red. \textbf{Bottom}: Residuals from the best fit.}
\label{fig:WASP-4}
\end{figure*}

\begin{table*}
	\centering
	\caption{\texttt{allesfitter} model priors and fitted values. Planetary System properties for the transiting system WASP-4 (TIC 402026209), as shown in \autoref{fig:WASP-4}.}
    \begin{tabular}{lcccc}
    \hline
	\textbf{Parameter}&\textbf{Initial Guess}&\textbf{Prior}&\textbf{Fitted value CMOS}&\textbf{Fitted value CCD} \\
    \hline
    $R_p / R_{\star}$            & \FITNTrb          & $\mathcal{N}\left(0.15201,0.00040\right)$         & \CMOSNTrb	    & \CCDNTrb   \\
	$(R_\star + R_b) / a_b$      & \FITNTrsuma       & $\mathcal{N}\left(0.211339,0.002017\right)$     & \CMOSNTrsuma	& \CCDNTrsuma   \\
    $\cos{i_b}$                  & \FITNTinc         & $\mathcal{N}\left(0.016405,0.014657\right)$     & \CMOSNTinc	    & \CCDNTinc   \\
    Tc (BJD)		             & \FITNTtc	         & $\mathcal{N}\left(2455804.515752,0.000019\right)$ & \CMOSNTtc	    & \CCDNTtc	\\
    Orbital Period (days)	  	 & \FITNTperiod      & $\mathcal{N}\left(1.338231466,0.000000023\right)$   & \CMOSNTperiod	& \CCDNTperiod	\\
	$q_1$                        & \FITNTqone        & $\mathcal{N}\left(0.385,0.012\right)$           & \CMOSNTqone	& \CCDNTqone   \\
	$q_2$                        & \FITNTqtwo        & $\mathcal{N}\left(0.400,0.087\right)$           & \CMOSNTqtwo	& \CCDNTqtwo   \\
    \hline
    \end{tabular}
    \label{tab:planetary-wasp-4}
\end{table*}

\section{Discussion}
\label{sec:discussion}

\subsection{Fractional Noise}
\label{sec:noises_discussion}
The RMS precision for lightcurves generated from the CMOS images agree well with our noise model for instrumentation, sky background, and atmospheric scintillation noise (see right-hand panels in \autoref{fig:noise_dark_bright}). The CCD data are also in good agreement with the noise model (see left-hand panels in \autoref{fig:noise_dark_bright}).

We note that for the CMOS camera, the purely instrumental noise factors (dark current and read noise) are never a significant noise source, even for faint stars on nights with no moon. For the CCD camera, however, the read noise is at the same level as the sky background during no-moon conditions for stars fainter than T>$13$\,mag. This difference is primarily due to the fact that the CCD camera has much higher read noise than the CMOS camera (see \autoref{tab:specs_results}). This difference in read noise arises because CMOS pixels are read out in parallel, allowing the amplifiers to process signals at significantly lower frequencies. In contrast, CCD cameras have fewer amplifiers, requiring them to operate at higher frequencies, which leads to larger read noise \citep{1485947, 628823, Ioannis24}.
 
Although the dark current and amplifier glow contributions for the CMOS image sensors are much higher than the CCD image sensors \citep{2021RMxAC..53..190K, 2021RAA....21..268Q, Ioannis24}, the dark current is not a significant noise source even for the CMOS camera for this photometry. For the CMOS camera, observations of bright stars under no-moon conditions, we are either limited by atmospheric scintillation or photon shot noise from the target star. For fainter stars, photometric precision is limited by sky background noise, primarily due to the large pixel scale (4\,arcsec pix$^{-1}$) and the point-spread function (FWHM=1.7\,pix). This limitation can be mitigated by reducing the photometric aperture size, which helps exclude excess background signal and improves signal-to-noise for fainter stars. We note that if there was a much finer pixel scale and narrower point-spread-function, then the sky contribution would be much less and therefore the dark current would start to become the dominant noise contribution for objects fainter than approximately T$=15$\,mag. During full-moon conditions, the increased sky background means that all stars with T$ >10.5$\,mag will be limited by sky noise. This is also the case for the CCD camera.

We observed a distinct increase in the RMS for the very brightest stars (T$ <9.5$\,mag) for the CMOS camera, which was less pronounced for the CCD camera (see \autoref{fig:noise_dark_bright}). This effect is due to the saturation of the pixel's capacity of the CMOS camera and it is avoided by reducing the exposure time. The CMOS architecture is equipped with transistors within each pixel that amplify and convert charge to voltage. This process involves transferring charge from the photodiode by applying a specific voltage to the gate, which serves as an anti-blooming mechanism, preventing the escape of charge to adjacent pixels \citep{stefanov2022cmos, Ioannis24}. The overall result of this is that charge is not conserved for saturated pixels, hence any relative photometry of a star with a saturated pixel will have a higher RMS proportional to the degree of saturation. A saturated star will not exhibit the same flux–airmass correlation as unsaturated stars, because its recorded counts are clipped by the pixel full well capacity rather than tracking the true incident flux. As a result, standard airmass-based detrending becomes ineffective for such targets, and applying it can introduce additional scatter into the light curve. This is evidenced by the increasing RMS with \tess\ magnitude for stars with T$ > 9$\,mag in \autoref{fig:noise_dark_bright}. Although the CCD camera also saturates, the CCD photometry is more precise for these very bright stars. This is due to the sensor architecture of CCD cameras, whereby if a pixel exceeds its full well depth, the excess charge spreads to neighbouring pixels \citep{1987OptEn..26..692J}. This results in bleeding or smearing artifacts, typically along the direction of charge transfer of the readout \citep{1987OptEn..26..692J}. For very bright stars, these effects were evident in the CCD frames. In some cases, saturated pixels leaking charge into nearby pixels within the aperture can still be used for photometry, as demonstrated by \cite{2016MNRAS.455L..36P}. 

Both cameras demonstrate performance that aligns well with the noise model under both no-moon and full-moon sky conditions, achieving scintillation-limited performance for the brightest stars. 

\begin{figure}
\includegraphics[width=\columnwidth]{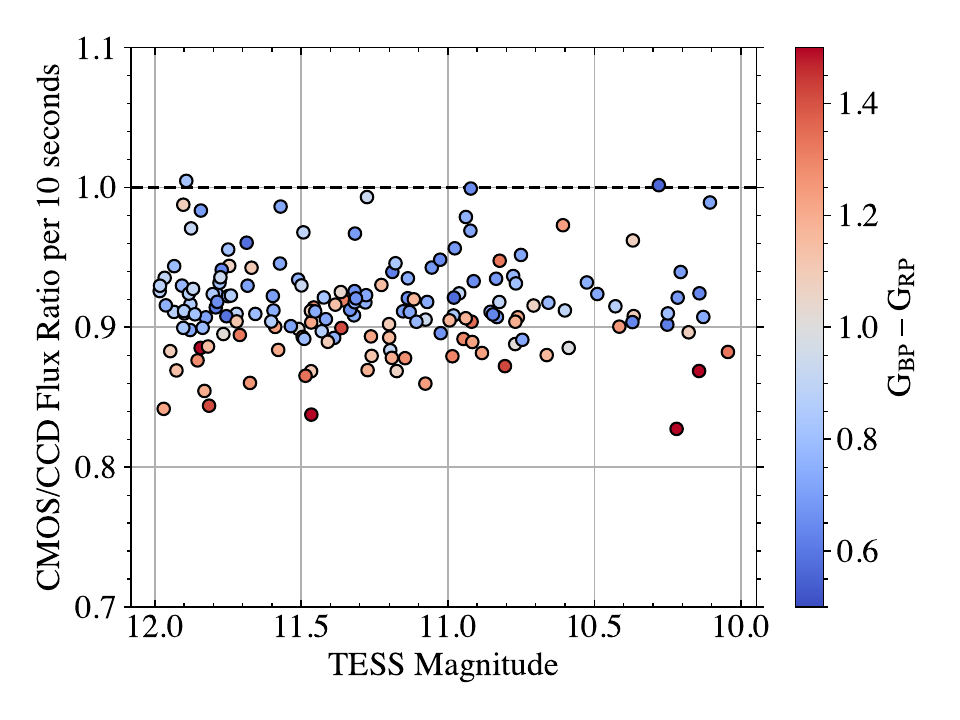}
\caption{The CMOS/CCD flux ratio for a set of stars from the night of 2024 July 5. Flux ratios are represented by circles, colour-coded according to the catalogued \textit{Gaia} DR3 $\mathrm{G_{BP}}-\mathrm{G_{RP}}$ colour. The dashed line represents the ratio 1 which is the case where the detected flux is equal for both cameras.}
\label{fig:ratio_flux_10}
\end{figure}

Upon analysing the noise ratio comparison shown in \autoref{fig:rms_ratio}, we find that the CCD camera exhibits lower noise for bright stars ($10 < \mathrm{T} < 12$). In this regime, the photon shot noise from the target stars dominates, and the CCD receives more signal than the CMOS for the same 10-second exposure. This behaviour arises because the NGTS filter is optimised for the CCD’s quantum efficiency response. In contrast, the CMOS sensor, which is more sensitive to bluer wavelengths, operates with reduced efficiency since the NGTS filter cuts off much of its most sensitive spectral region (see \autoref{fig:filters_ratio}). To demonstrate this, we plot the ratio of the average fluxes of catalog stars as a function of stellar effective temperature, as shown in \autoref{fig:ratio_flux_10}. In this example, we consider only the exposure times and do not account for duty cycles (see \autoref{fig:ratio_flux} for the full comparison). We find that the CCD consistently records higher flux values than the CMOS camera, with the mean ratio of CMOS/CCD flux being approximately 0.9. We also see the expected colour dependence, with bluer stars showing a higher flux ratio due to the fact that the CMOS quantum efficiency is more blue sensitive (see \autoref{fig:filters_ratio}).

A clear transition is observed beyond $\mathrm{T} > 12$\,mag, and the CMOS camera achieves better precision. A similar pattern is seen under full-moon conditions (lower panel of \autoref{fig:rms_ratio}), although the ratio distribution peaks at approximately 0.95. The transition is likely due to the fact that the CCD camera is more sensitive to the sky background signal compared to the CMOS camera (see \autoref{fig:cmos_ccd_frames} and the discussion of the airglow later in Section~\ref{sec:sky_efficiency}). At the bright end ($\mathrm{T} < 9$\,mag), direct comparison becomes impractical due to CMOS saturation. It should be noted that the comparisons in \autoref{fig:noise_dark_bright} and \autoref{fig:rms_ratio} are made per exposure, without accounting for cadence.

\subsection{Sky Background}
\label{sec:sky_efficiency}
As shown in \autoref{fig:noise_dark_bright}, sky-background is the dominant noise source for both our CMOS and CCD photometry for stars with $\mathrm{T} > 13$\,mag in no-moon conditions and $\mathrm{T} > 11$\,mag in full-moon conditions (using our chosen source aperture size). The CCD camera is deep-depleted, with its quantum efficiency peaking at red wavelengths (see \autoref{fig:filters_ratio}). This makes the CCD camera more sensitive to the strong emission lines redwards of 600\,nm, such as those from atomic oxygen and OH produced in the mesosphere \citep{1993PASP..105..940M}. This phenomenon, commonly referred to as airglow \citep{2012A&A...543A..92N}, remains detectable despite the NGTS having a cut-off at 890\,nm. In contrast, the CMOS camera is less sensitive to red wavelengths and therefore less sensitive to airglow. This effect is evident when comparing simultaneous images of the sky background with the CMOS and CCD camera. For example, the target pixel frame in \autoref{fig:cmos_ccd_frames} shows the CCD camera to have a sky background count of 4.02\,e$^-$\,s$^{-1}$\,arcsec$^{-2}$, where the CMOS camera's sky background count is only 2.52\,e$^-$\,s$^{-1}$\,arcsec$^{-2}$.  

\begin{figure*}
\includegraphics[width=\textwidth]{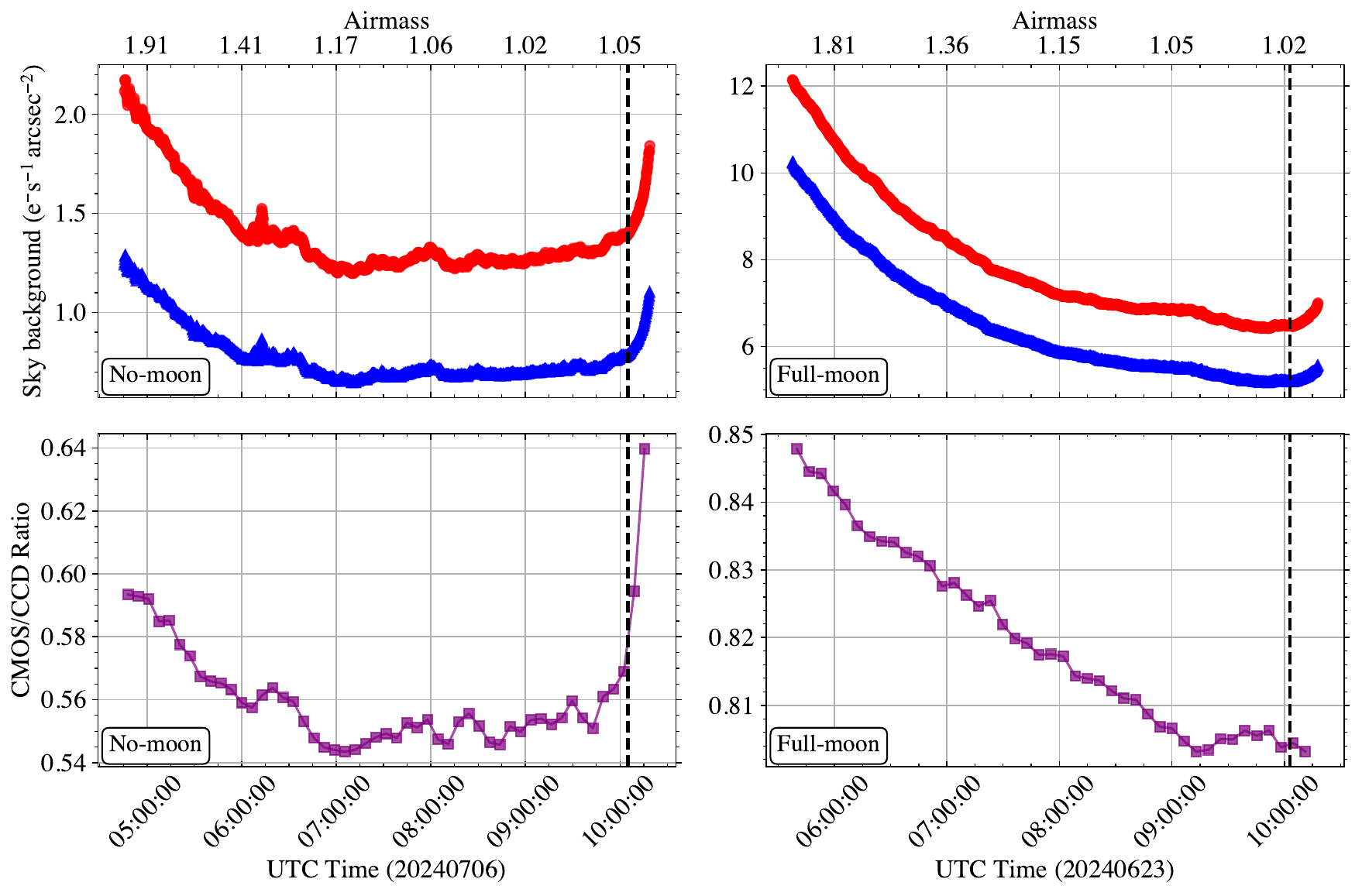}
\caption{\textbf{Top:} Sky background flux (e$^-$\,pixel$^{-1}$\,arcsec$^{-2}$) as a function of time and airmass from the night of 2024 July 5 (\textbf{left}) and the night of 2024 June 22 (\textbf{right}) for the CMOS (blue triangles) and CCD (red circles). The dashed vertical line at UT10:05:00 (\textbf{left}) and UT10:03:00 (\textbf{right}) indicates the start of astronomical twilight. \textbf{Bottom:} The CMOS/CCD ratio of the fluxes from the top panel as a function of time.}
\label{fig:ratio_sky}
\end{figure*}

To examine this effect further, we measured the sky background for both cameras over a 6-hour observation on the no-moon night of 2024 July 5 and on the full-moon night of 2024 June 22.  We used the counts from the sky background map. The digital units were converted to electrons and normalised by the exposure time and sky pixel area in arcsec$^{-2}$. The sky background flux in units of e$^-$\,pixel$^{-1}$\,arcsec$^{-2}$ is shown in \autoref{fig:ratio_sky} as a function of time. The measurements for the no-moon dataset show that although both cameras have a similar trend with time across the observation block, the sky background flux for the CMOS camera is consistently lower than the CCD camera by a factor of approximately 0.5 (see lower panel of \autoref{fig:ratio_sky}). In contrast, the measurements for the full-moon show the contribution of the moon light which follows the same trend for both cameras while the sky background has increased significantly. The CMOS camera is consistently lower than the CCD camera however the ratio has been increased by a factor of approximately 0.8 (see lower panel of \autoref{fig:ratio_sky}). Given the only difference is the quantum efficiency of the cameras, the higher sky background for the CCD camera can be attributed to the fact that the sky is brighter in the red wavelengths due to the night sky emission, and therefore the sky background is higher for the more red-sensitive CCD camera.

We note that the overall trend seen for both cameras in \autoref{fig:ratio_sky} is due to the fact that the airmass is high (z $= 1.91$) at the start of the observations. That increase in amplitude just before the end of the observations is primarily due to the fact that we are observing at the astronomical twilight. 


\subsection{Camera Efficiency}
\label{sec:camera_efficiency}
We compare the amount of flux collected per unit time for each camera, which takes into account the sum of the exposure times (10\,s) and dead times (3\,s for the CCD camera and 0.042\,s for the CMOS camera). The flux collected is also dependent on the quantum efficiency of each camera (see \autoref{fig:filters_ratio}), the NGTS bandpass (520-890\,nm) and the stellar effective temperature. We assume that the throughput of each NGTS telescope is identical.  

\begin{figure}
\includegraphics[width=\columnwidth]{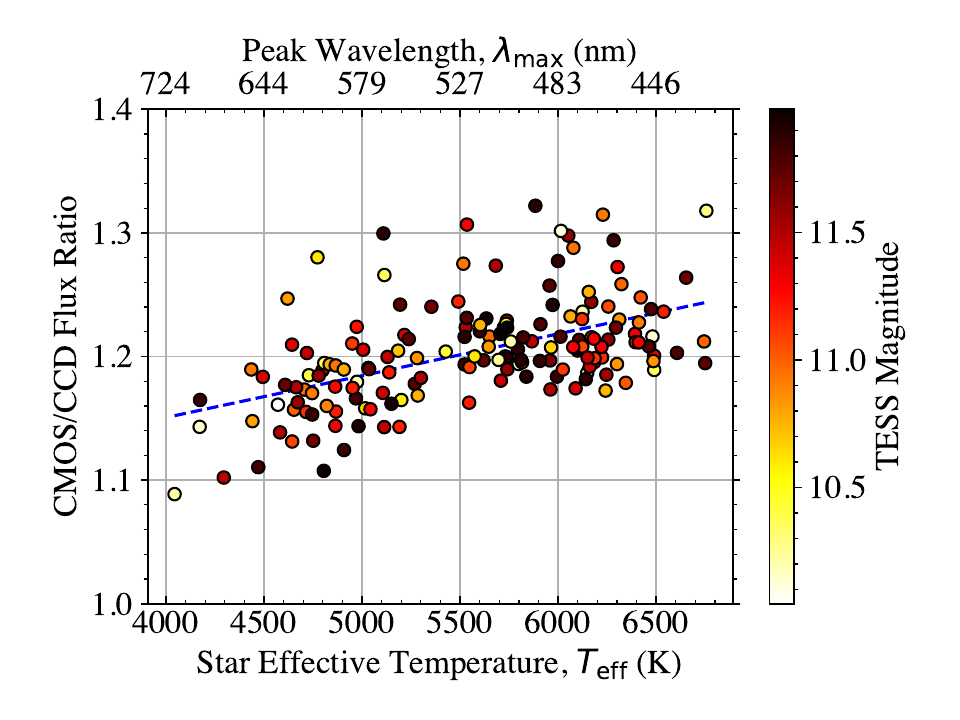}
\caption{Ratio of the flux for bright stars ($\mathrm{T} < 12$\,mag), plotted against stellar effective temperature and corresponding peak wavelength, assuming each star behaves as a blackbody, as measured by the CMOS and CCD cameras from the night of 2024 July 5. Flux ratios are represented by circles, colour-coded according to TESS magnitude. The blue dashed line is best-fit linear model the data.}
\label{fig:ratio_flux}
\end{figure}

We focused only on stars brighter than $\mathrm{T} < 12$\,mag. Flux measurements for each star were obtained by aperture photometry using a constant physical photometric radius for the CCD and CMOS images (55\,$\unit{\um}$ - 20\,arcsec). The duration is determined by the combined exposure and readout times of the respective camera. To evaluate the performance, we computed the ratio of flux in electrons per unit time between the two cameras and compare it to the stellar effective temperature of each observed star, as shown in \autoref{fig:ratio_flux}. 

Overall, the flux ratio between the CMOS and CCD cameras is always greater than unity, which shows that the CMOS camera collects more electrons per unit elapsed time than the CCD camera. This is primarily due to the fact that the CMOS camera reads out approximately 70 times faster than the CCD camera when including the dead times. We find a trend in the flux ratio with the effective temperature of the star derived from the TIC8 \citep{2018AJ....156..102S} catalogue (see \autoref{fig:ratio_flux}). This trend is due to the fact that the CMOS camera is more blue-sensitive in its quantum efficiency compared to the CCD camera (see \autoref{fig:filters_ratio}). This means that the CMOS camera performs even better than the CCD camera for the bluest (hottest) stars. We do not detect any trend in the flux ratio with respect to the magnitude of the stars.  

\begin{figure}
\includegraphics[width=\columnwidth]{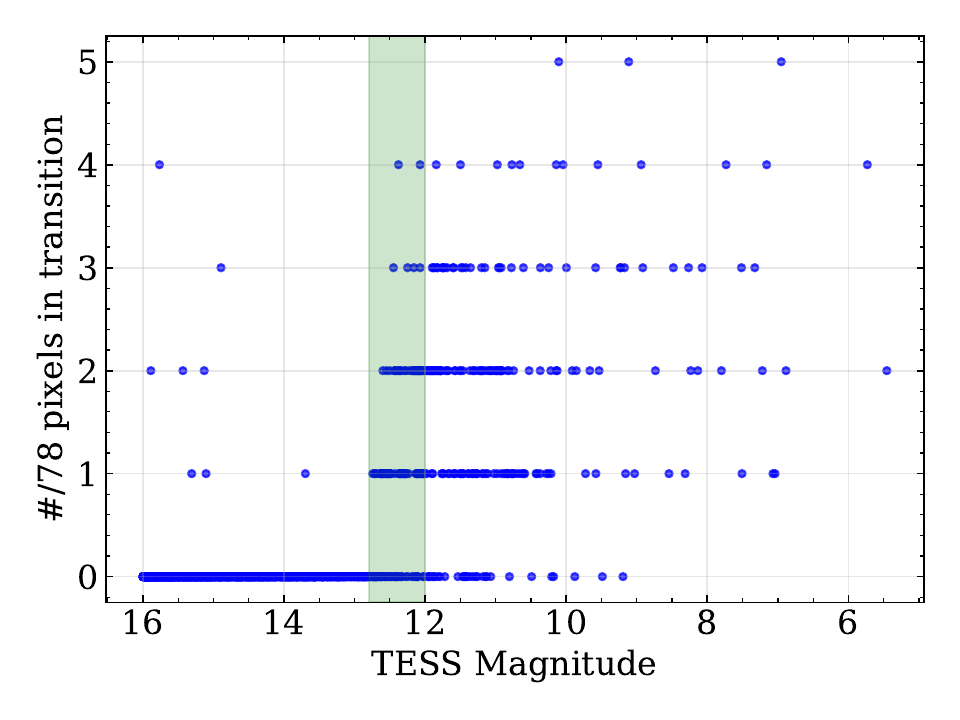}
\caption{Number of transition pixels (1600-2000\,ADU) inside the photometric aperture ($\sim$78 pixels aperture) for a selection of stars observed with the CMOS camera on the night of 2024 July 5 (new moon conditions).
The green shaded area represents the magnitude range in which transition pixels have the maximum value within the photometric aperture.}
\label{fig:transition_bins}
\end{figure}

The Marana CMOS camera achieves an extended dynamic range through its HDR mode. In our previous work \citep{Ioannis24}, we demonstrated that a transition between the High Gain and Low Gain channels occurs at approximately $\sim$1800\,ADU. Pixels with counts in this regime may be processed by either gain channel, thereby introducing additional noise \citep[see Figure 9]{Ioannis24}. We investigated which stars are affected by such transition pixels within their photometric apertures. We defined the transition region to be 1600–2000\,ADU. For each star, we measured the number of pixels inside the aperture whose values fall within this range, and grouped stars into magnitude bins. This analysis is illustrated in \autoref{fig:transition_bins}.

Additionally, we identified stars for which the maximum pixel value inside the aperture, corresponding to the point spread function peak and contributing most significantly to the total flux, lies within the transition region. These stars were found to be in the range $12 < \mathrm{T} < 12.8$, as indicated by the green shaded region in \autoref{fig:transition_bins}. The number of affected stars per magnitude bin, together with the average number of transition pixels per aperture for each star, is presented in \autoref{tab:transition}. We also analysed the final relative photometry of stars whose point spread functions' contain transition pixels, and compared them with stars whose four highest pixel values do not include any transition pixels.

We find that for faint stars (T$>$13) there are no transition pixels in the photometric aperture.  For stars brighter than T$\sim$12 there are typically just a few transition pixels, and these pixels are not significantly contributing to the total flux measured for the star.  Therefore in both these instances it is unlikely our photometric measurements will be affected by the transition feature of the CMOS detector.  The magnitude $12 < \mathrm{T} < 12.8$ is at highest risk of the photometry being affected by the CMOS transition from High Gain to Low Gain channels.  We do not find a noticeable increase in the of RMS noise of these light curves. However it would be advisable to ensure prime targets of photometric monitoring are not in this region by selecting an appropriate exposure time for observations.

\subsection{Red Noise}
\label{sec:timescale_discussion}
Red-noise is a critical limitation for many applications of bright star photometry - e.g. for detecting exoplanet transits \citep{2006MNRAS.373..231P}.
Our analysis of the photometric precision for the CMOS and CCD cameras at different timescales (see Section~\ref{sec:timescale_results} and \autoref{fig:rms_timescale}) shows that the noise for both cameras is largely Gaussian (white) noise.  This is as expected from previous analysis of NGTS bright star photometry \citep{2017PASP..129b5002M, 2018MNRAS.475.4476W, 2020MNRAS.494.5872B, 2022SPIE12191E..1AB, 2022MNRAS.509.6111O}.

\section{Conclusions}
\label{sec:conclusion}
The Marana CMOS camera was studied extensively on-sky at the Paranal Observatory using an NGTS telescope to test its suitability for high-precision photometry of exoplanet transits. We conducted simultaneous observations using a CCD camera to ensure exactly the same observing conditions and the same targets. The common exposure time was selected to be 10\,s.  We made sure we accounted for the difference in pixel size between the two cameras (13.5 vs 11\,$\unit{\um}$ for the CCD and CMOS camera respectively). We also considered the differences in the quantum efficiency when reaching our conclusions between the two cameras (\autoref{fig:filters_ratio}).

The CMOS camera delivered photometry at the precision expected from our noise model, which was scintillation-limited for bright stars ($\mathrm{T} < 11$\,mag), and sky background-limited for fainter stars ($\mathrm{T} > 13$\,mag). In contrast to the CCD image sensor, the CMOS image sensor was never limited by readout noise. The CMOS camera's dark current noise is much higher than the CCD camera, but this was never a dominant source of noise for our photometry. We see minimal red noise contributions in the photometric precision of the CMOS camera, indicating there are no instrumental red-noise limitations using the CMOS camera for bright star photometry.

The fast readout of the CMOS camera (42\,ms) enables a much higher duty cycle than the CCD camera. This effectively enables more flux to be collected over any given observing duration, and hence a better photometric precision. This is even true despite the slightly reduced quantum efficiency of the CMOS image sensor compared with the CCD image sensor, and the fact that the NGTS filter has been optimised to the CCD camera's quantum efficiency response.
The transiting exoplanet lightcurves obtained using the CMOS camera are very similar to those obtained with the CCD camera. This result aligns well with other on-sky tests in the literature that show excellent photometric performance for CMOS cameras \citep{2021RMxAC..53..190K, alarcon2023scientific}.

In conclusion, we find that the Marana CMOS camera is suitable for high-precision time-series photometry. The fast readout speeds and low read noise and shutterless operation often provide a significant advantage over a traditional CCD camera. For applications requiring continuous imaging of rapidly varying astronomical targets, such as satellite tracking and mapping \citep{blake2023exploring, cooke2023simulated, 2025arXiv250212324A}, or near-Earth asteroid monitoring \citep{larson2018applications}, CMOS image sensors offer distinct benefits. Furthermore, image degradation caused by saturated stars is mitigated in CMOS cameras due to their anti-blooming pixel architecture, reducing the risk of contamination to nearby targets of interest. As CMOS sensor technology continues to advance, it is likely that these cameras will replace CCDs in most astronomical time-series photometry applications.

\section*{Acknowledgements}
This is a project based on the collaboration between the University of Warwick and Andor Technology Ltd under the STFC Industrial CASE (Cooperative Awards in Science and Technology) studentship. IA acknowledges the support of the UK Science and Technology Facilities Council (STFC) under the CASE Industry scheme (ST/W005077/1 Project title: Precision Photometry with the new generation of fast readout Scientific CMOS Camera). We acknowledge capital funding from the STFC Early Technology Development Capital Funding (ST/W005719/1 Project title:
Discovering New Worlds with sCMOS Cameras). This work is based on data collected under the NGTS project at the ESO La Silla Paranal Observatory. The NGTS facility is operated by a consortium institutes with support from the UK Science and Technology Facilities Council (STFC) under projects ST/M001962/1, ST/S002642/1 and ST/W003163/1. JSJ gratefully acknowledges support by FONDECYT grant 1240738 and from the ANID BASAL project FB210003.
ML acknowledges support of the Swiss National Science Foundation under grant number PCEFP2\_194576. The contribution of ML has been carried out within the framework of the NCCR PlanetS supported by the Swiss National Science Foundation under grants 51NF40\_182901 and 51NF40\_205606. S.M.O. is supported by a UK Science and Technology Facilities Council (STFC) Studentship (ST/W507751/1). 
\section*{Data Availability}
The software used is publicly available on the GitHub repository: \url{https://github.com/AperpieGG/ngcmos}. The full photometric dataset from \ngts\ will be available upon request to the author.


\bibliographystyle{rasti}
\bibliography{marana} 




\appendix
\section{Appendix}
\label{sec:appendix}
\begin{figure*}
\centering
\includegraphics[width=\textwidth]{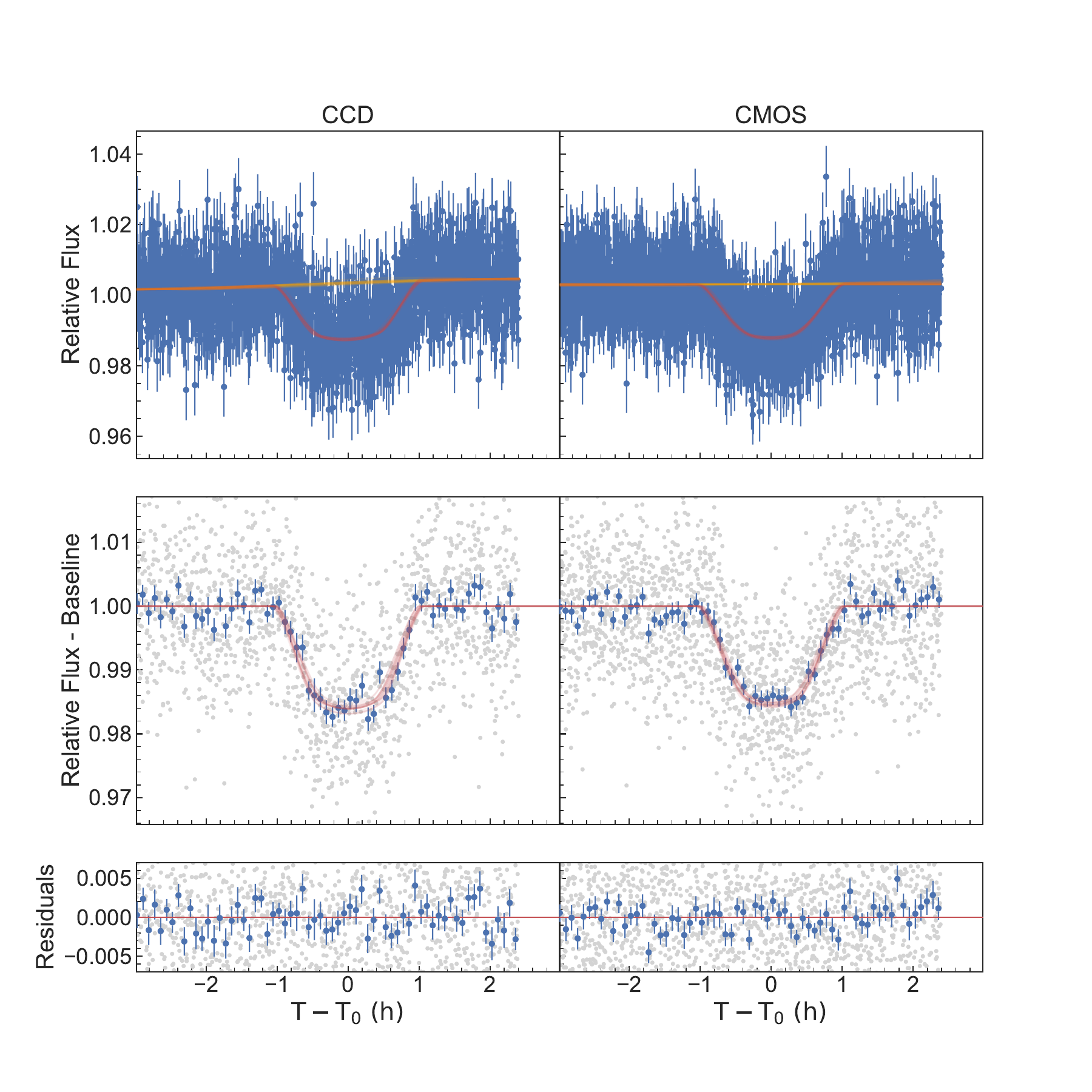}
\caption{Same as in \autoref{fig:WASP-4}. Normalized lightcurves for the transiting system of TOI-905 (TIC 261867566) from NGTS on 2024 July 13.}
\label{fig:toi-905}
\end{figure*}

\begin{table*}
	\centering
	\caption{\texttt{allesfitter} model priors. Planetary System properties for the transiting system TOI-905 (TIC 261867566) from NGTS on 2024 July 13.}
    \begin{tabular}{lcccc}
    \hline
	\textbf{Parameter}&\textbf{Initial Guess}&\textbf{Prior}&\textbf{Fitted value CMOS}&\textbf{Fitted value CCD} \\
    \hline
    $R_p / R_{\star}$            & \FITNArb          & $\mathcal{N}\left(0.1311,0.0012\right)$         & \CMOSNArb	    & \CCDNArb   \\
	$(R_\star + R_b) / a_b$      & \FITNArsuma       & $\mathcal{N}\left(0.102827,0.010844\right)$     & \CMOSNArsuma	& \CCDNArsuma   \\
    $\cos{i_b}$                  & \FITNAinc         & $\mathcal{N}\left(0.075327,0.004524\right)$     & \CMOSNAinc	    & \CCDNAinc   \\
    Tc (BJD)		             & \FITNAtc	         & $\mathcal{N}\left(2460505.6096463,0.0004719\right)$ & \CMOSNAtc	    & \CCDNAtc	\\
    Orbital Period (days)	  	 & \FITNAperiod      & $\mathcal{N}\left(3.7395671,0.0000013\right)$   & \CMOSNAperiod	& \CCDNAperiod	\\
	$q_1$                        & \FITNAqone        & $\mathcal{N}\left(0.387,0.015\right)$           & \CMOSNAqone	& \CCDNAqone   \\
	$q_2$                        & \FITNAqtwo        & $\mathcal{N}\left(0.408,0.099\right)$           & \CMOSNAqtwo	& \CCDNAqtwo   \\
    \hline
    \end{tabular}
    \label{tab:planetary-TOI-905_0713}
\end{table*}

\begin{figure*}
\centering
\includegraphics[width=\textwidth]{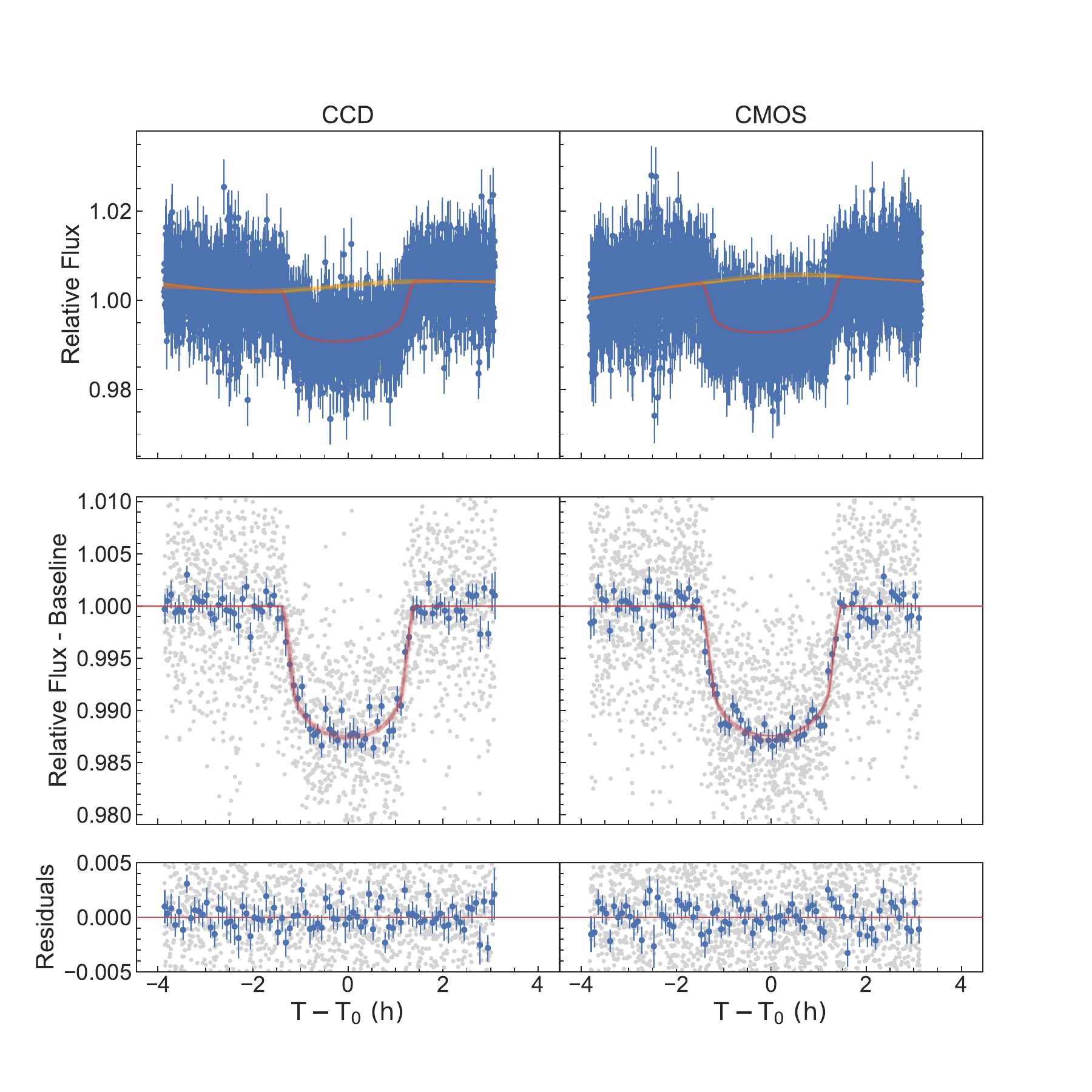}
\caption{Same as in \autoref{fig:WASP-4}. Normalized lightcurves for the transiting system of WASP-95 (TIC 144065872) from NGTS on 2024 July 14.}
\label{fig:wasp-95}
\end{figure*}

\begin{table*}
	\centering
	\caption{\texttt{allesfitter} model priors. Planetary System properties for the transiting system WASP-95 (TIC 144065872) from NGTS on 2024 July 14.}
    \begin{tabular}{lcccc}
    \hline
	\textbf{Parameter}&\textbf{Initial Guess}&\textbf{Prior}&\textbf{Fitted value CMOS}&\textbf{Fitted value CCD} \\
    \hline
    $R_p / R_{\star}$            & \FITNBrb          & $\mathcal{N}\left(0.1025,0.0005\right)$         & \CMOSNBrb	    & \CCDNBrb   \\
	$(R_\star + R_b) / a_b$      & \FITNBrsuma       & $\mathcal{N}\left(0.172266,0.005922\right)$     & \CMOSNBrsuma	& \CCDNBrsuma   \\
    $\cos{i_b}$                  & \FITNBinc         & $\mathcal{N}\left(0.027922,0.036611\right)$     & \CMOSNBinc	    & \CCDNBinc   \\
    Tc (BJD)		             & \FITNBtc	         & $\mathcal{N}\left(2458553.711158,0.000088\right)$ & \CMOSNBtc	    & \CCDNBtc	\\
    Orbital Period (days)	  	 & \FITNBperiod      & $\mathcal{N}\left(2.1846730,0.0000014\right)$   & \CMOSNBperiod	& \CCDNBperiod	\\
	$q_1$                        & \FITNBqone        & $\mathcal{N}\left(0.357,0.012\right)$           & \CMOSNBqone	& \CCDNBqone   \\
	$q_2$                        & \FITNBqtwo        & $\mathcal{N}\left(0.393,0.089\right)$           & \CMOSNBqtwo	& \CCDNBqtwo   \\
    \hline
    \end{tabular}
    \label{tab:planetary-WASP-95_0714}
\end{table*}


\begin{figure*}
\centering
\includegraphics[width=\textwidth]{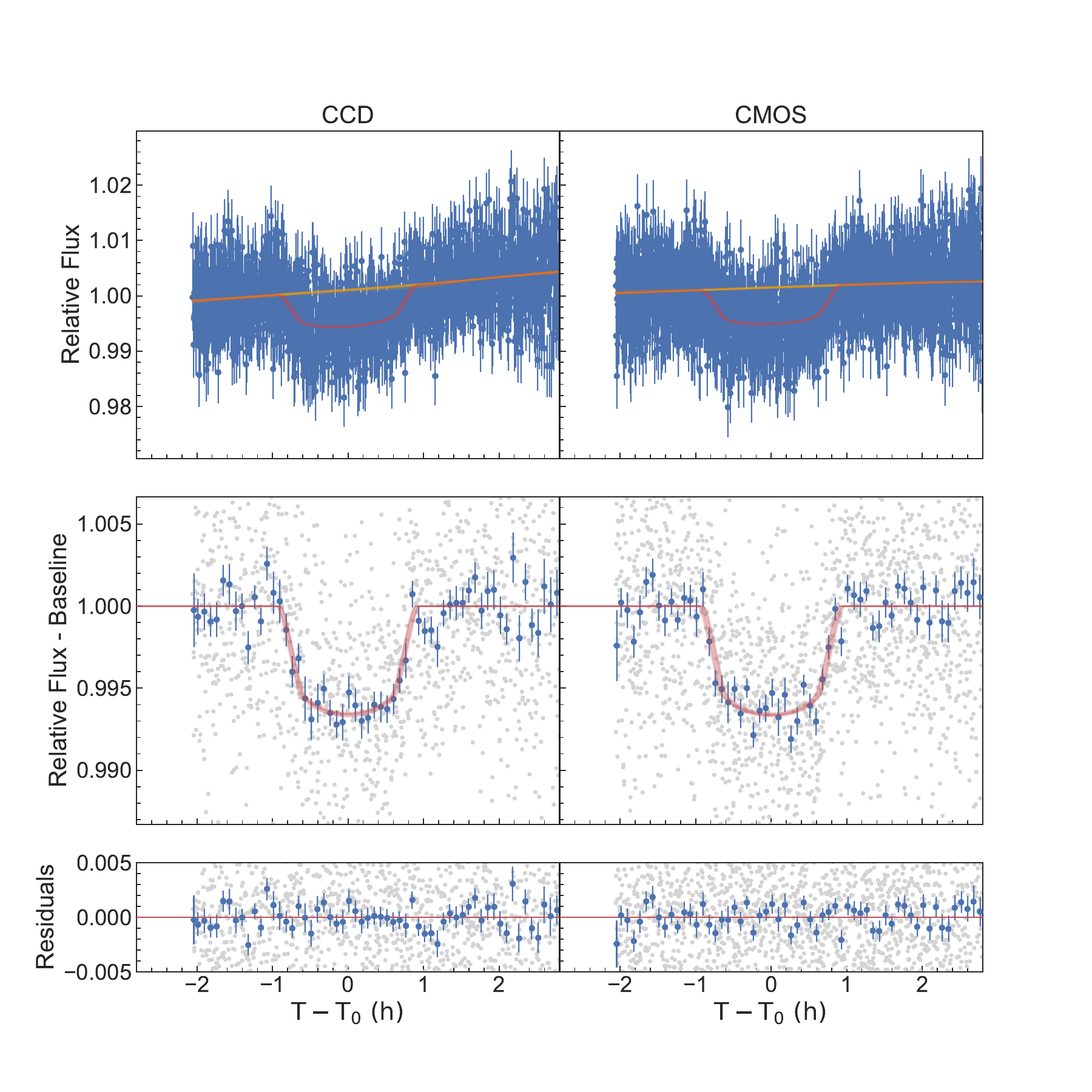}
\caption{Same as in \autoref{fig:WASP-4}. Normalized lightcurves for the transiting system of TOI-2109 (TIC 392476080) from NGTS on 2024 July 17.}
\end{figure*}

\begin{table*}
	\centering
	\caption{\texttt{allesfitter} model priors and fitted values. Planetary System properties for the transiting system TOI-2109 (TIC 392476080) from NGTS on 2024 July 17.}
    \begin{tabular}{lcccc}
    \hline
	\textbf{Parameter}&\textbf{Initial Guess}&\textbf{Prior}&\textbf{Fitted value CMOS}&\textbf{Fitted value CCD} \\
    \hline
    $R_p / R_{\star}$            & \FITNCrb          & $\mathcal{N}\left(0.08155,0.00022\right)$         & \CMOSNCrb	    & \CCDNCrb   \\
	$(R_\star + R_b) / a_b$      & \FITNCrsuma       & $\mathcal{N}\left(0.476874,0.044155\right)$     & \CMOSNCrsuma	& \CCDNCrsuma   \\
    $\cos{i_b}$                  & \FITNCinc         & $\mathcal{N}\left(0.329855,0.006089\right)$     & \CMOSNCinc	    & \CCDNCinc   \\
    Tc (BJD)		             & \FITNCtc	         & $\mathcal{N}\left(2459378.455370,0.000059\right)$ & \CMOSNCtc	    & \CCDNCtc	\\
    Orbital Period (days)	  	 & \FITNCperiod      & $\mathcal{N}\left(0.672474140,0.000000028\right)$   & \CMOSNCperiod	& \CCDNCperiod	\\
	$q_1$                        & \FITNCqone        & $\mathcal{N}\left(0.305,0.008\right)$           & \CMOSNCqone	& \CCDNCqone   \\
	$q_2$                        & \FITNCqtwo        & $\mathcal{N}\left(0.375,0.074\right)$           & \CMOSNCqtwo	& \CCDNCqtwo   \\
    \hline
    \end{tabular}
    \label{tab:planetary-TOI-2109_0717}
\end{table*}

\begin{figure*}
\centering
\includegraphics[width=\textwidth]{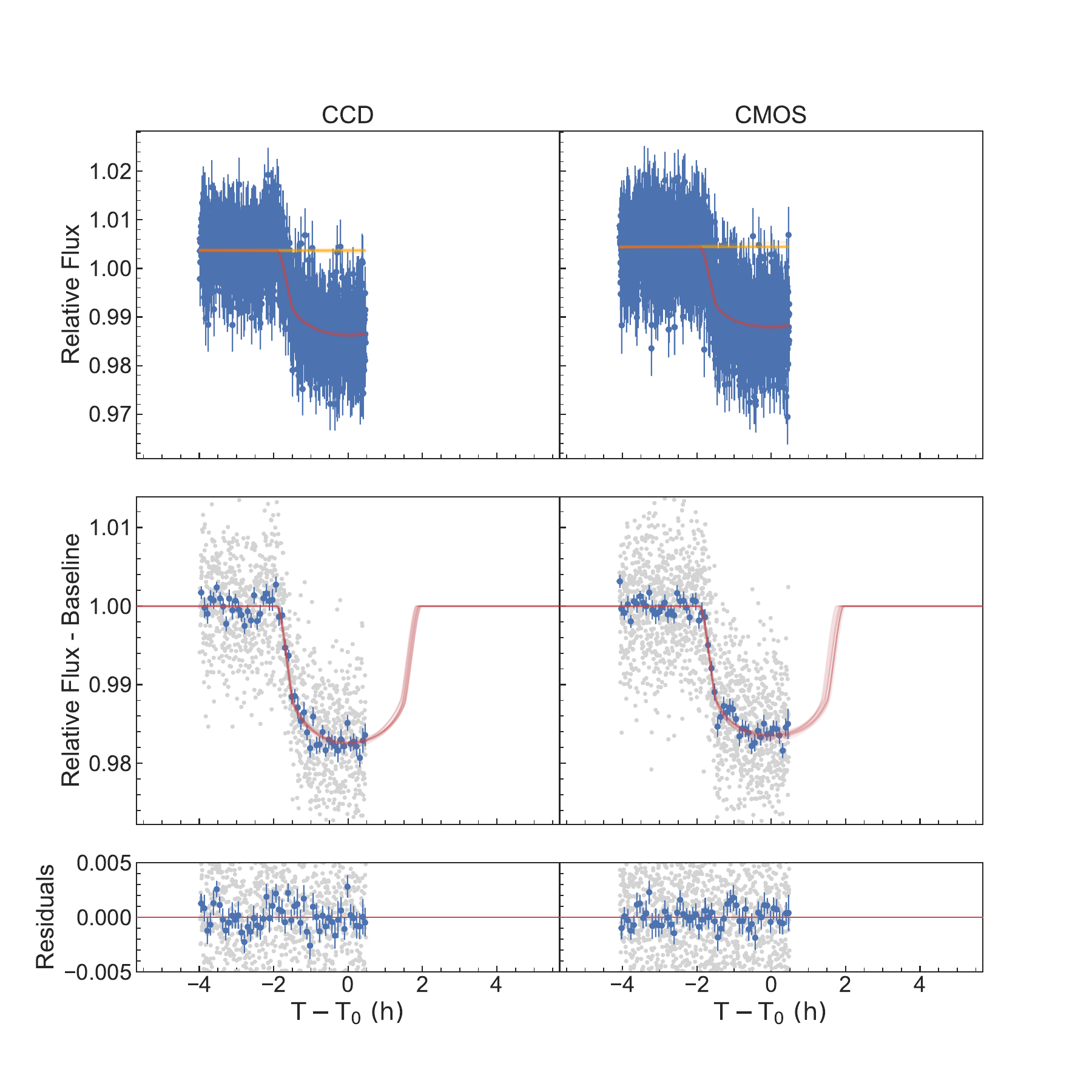}
\caption{Same as in \autoref{fig:WASP-4}. Normalized lightcurves for the transiting system of KELT-10 (TIC 269217040) from NGTS on 2024 August 1.}
\end{figure*}

\begin{table*}
	\centering
	\caption{\texttt{allesfitter} model priors and fitted values. Planetary System properties for the transiting system KELT-10 (TIC 269217040) from NGTS on 2024 August 1.}
    \begin{tabular}{lcccc}
    \hline
	\textbf{Parameter}&\textbf{Initial Guess}&\textbf{Prior}&\textbf{Fitted value CMOS}&\textbf{Fitted value CCD} \\
    \hline
    $R_p / R_{\star}$            & \FITNDrb          & $\mathcal{N}\left(0.1190,0.0014\right)$         & \CMOSNDrb	    & \CCDNDrb   \\
	$(R_\star + R_b) / a_b$      & \FITNDrsuma       & $\mathcal{N}\left(0.119807,0.004107\right)$     & \CMOSNDrsuma	& \CCDNDrsuma   \\
    $\cos{i_b}$                  & \FITNDinc         & $\mathcal{N}\left(0.024258,0.015002\right)$     & \CMOSNDinc	    & \CCDNDinc   \\
    Tc (BJD)		             & \FITNDtc	         & $\mathcal{N}\left(2457612.49947,0.00041\right)$ & \CMOSNDtc	    & \CCDNDtc	\\
    Orbital Period (days)	  	 & \FITNDperiod      & $\mathcal{N}\left(4.1662541,0.0000016\right)$   & \CMOSNDperiod	& \CCDNDperiod	\\
	$q_1$                        & \FITNDqone        & $\mathcal{N}\left(0.350,0.010\right)$           & \CMOSNDqone	& \CCDNDqone   \\
	$q_2$                        & \FITNDqtwo        & $\mathcal{N}\left(0.385,0.081\right)$           & \CMOSNDqtwo	& \CCDNDqtwo   \\
    \hline
    \end{tabular}
    \label{tab:planetary-KELT-10_0801}
\end{table*}

\begin{figure*}
\centering
\includegraphics[width=\textwidth]{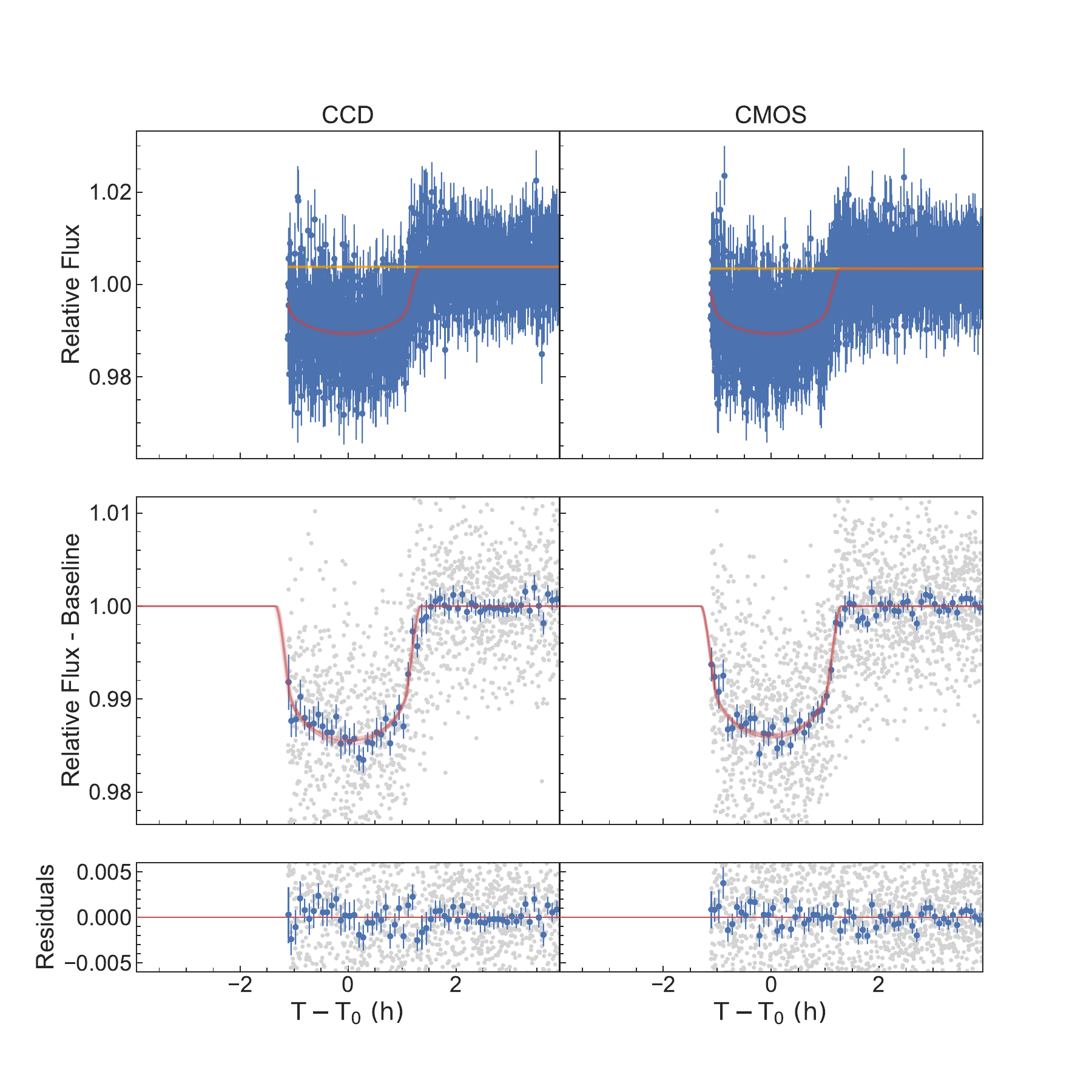}
\caption{Same as in \autoref{fig:WASP-4}. Normalized lightcurves for the transiting system of WASP-97 (TIC 230982885) from NGTS on 2024 August 04.}
\end{figure*}

\begin{table*}
	\centering
	\caption{\texttt{allesfitter} model priors and fitted values. Planetary System properties for the transiting system WASP-97 (TIC 230982885) from NGTS on 2024 August 04.}
    \begin{tabular}{lcccc}
    \hline
	\textbf{Parameter}&\textbf{Initial Guess}&\textbf{Prior}&\textbf{Fitted value CMOS}&\textbf{Fitted value CCD} \\
    \hline
    $R_p / R_{\star}$            & \FITNErb          & $\mathcal{N}\left(0.1091,0.0009\right)$         & \CMOSNErb	    & \CCDNErb   \\
	$(R_\star + R_b) / a_b$      & \FITNErsuma       & $\mathcal{N}\left(0.168300,0.003833\right)$     & \CMOSNErsuma	& \CCDNErsuma   \\
    $\cos{i_b}$                  & \FITNEinc         & $\mathcal{N}\left(0.034899,0.022665\right)$     & \CMOSNEinc	    & \CCDNEinc   \\
    Tc (BJD)		             & \FITNEtc	         & $\mathcal{N}\left(2458554.475352,0.000074\right)$ & \CMOSNEtc	    & \CCDNEtc	\\
    Orbital Period (days)	  	 & \FITNEperiod      & $\mathcal{N}\left(2.07275965,0.00000021\right)$   & \CMOSNEperiod	& \CCDNEperiod	\\
	$q_1$                        & \FITNEqone        & $\mathcal{N}\left(0.369,0.009\right)$           & \CMOSNEqone	& \CCDNEqone   \\
	$q_2$                        & \FITNEqtwo        & $\mathcal{N}\left(0.400,0.076\right)$           & \CMOSNEqtwo	& \CCDNEqtwo   \\
    \hline
    \end{tabular}
    \label{tab:planetary-WASP-97_0804}
\end{table*}

\begin{figure*}
\centering
\includegraphics[width=\textwidth]{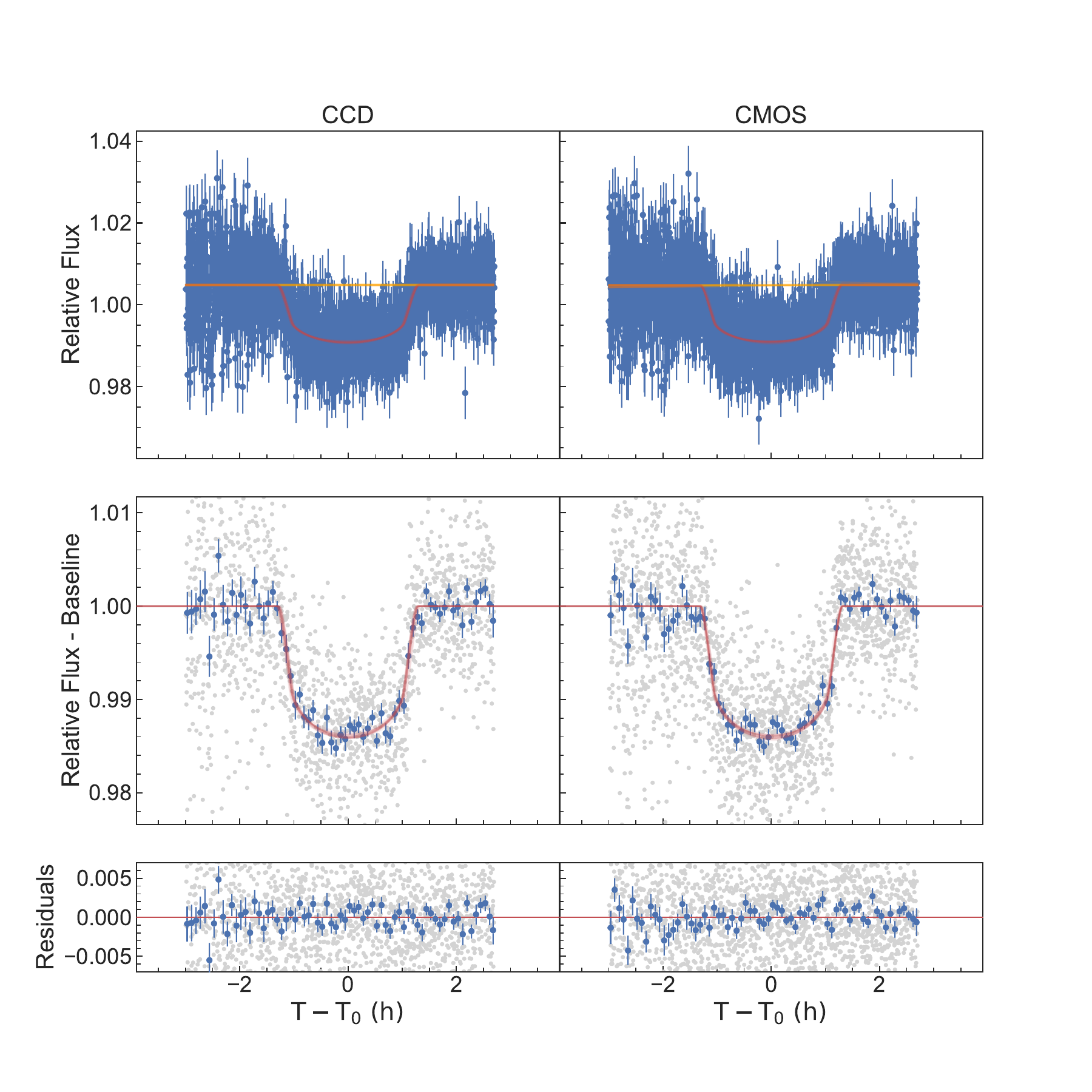}
\caption{Same as in \autoref{fig:WASP-4}. Normalized lightcurves for the transiting system of WASP-97 (TIC 230982885) from NGTS on 2024 August 06.}
\end{figure*}

\begin{table*}
	\centering
	\caption{\texttt{allesfitter} model priors and fitted values. Planetary System properties for the transiting system WASP-97 (TIC 230982885) from NGTS on 2024 August 06.}
    \begin{tabular}{lcccc}
    \hline
	\textbf{Parameter}&\textbf{Initial Guess}&\textbf{Prior}&\textbf{Fitted value CMOS}&\textbf{Fitted value CCD} \\
    \hline
    $R_p / R_{\star}$            & \FITNFrb          & $\mathcal{N}\left(0.1091,0.0009\right)$         & \CMOSNFrb	    & \CCDNFrb   \\
	$(R_\star + R_b) / a_b$      & \FITNFrsuma       & $\mathcal{N}\left(0.168300,0.003833\right)$     & \CMOSNFrsuma	& \CCDNFrsuma   \\
    $\cos{i_b}$                  & \FITNFinc         & $\mathcal{N}\left(0.034899,0.022665\right)$     & \CMOSNFinc	    & \CCDNFinc   \\
    Tc (BJD)		             & \FITNFtc	         & $\mathcal{N}\left(2458554.475352,0.000074\right)$ & \CMOSNFtc	    & \CCDNFtc	\\
    Orbital Period (days)	  	 & \FITNFperiod      & $\mathcal{N}\left(2.07275965,0.00000021\right)$   & \CMOSNFperiod	& \CCDNFperiod	\\
	$q_1$                        & \FITNFqone        & $\mathcal{N}\left(0.369,0.009\right)$           & \CMOSNFqone	& \CCDNFqone   \\
	$q_2$                        & \FITNFqtwo        & $\mathcal{N}\left(0.400,0.076\right)$           & \CMOSNFqtwo	& \CCDNFqtwo   \\
    \hline
    \end{tabular}
    \label{tab:planetary-WASP-97_0806}
\end{table*}

\begin{figure*}
\centering
\includegraphics[width=\textwidth]{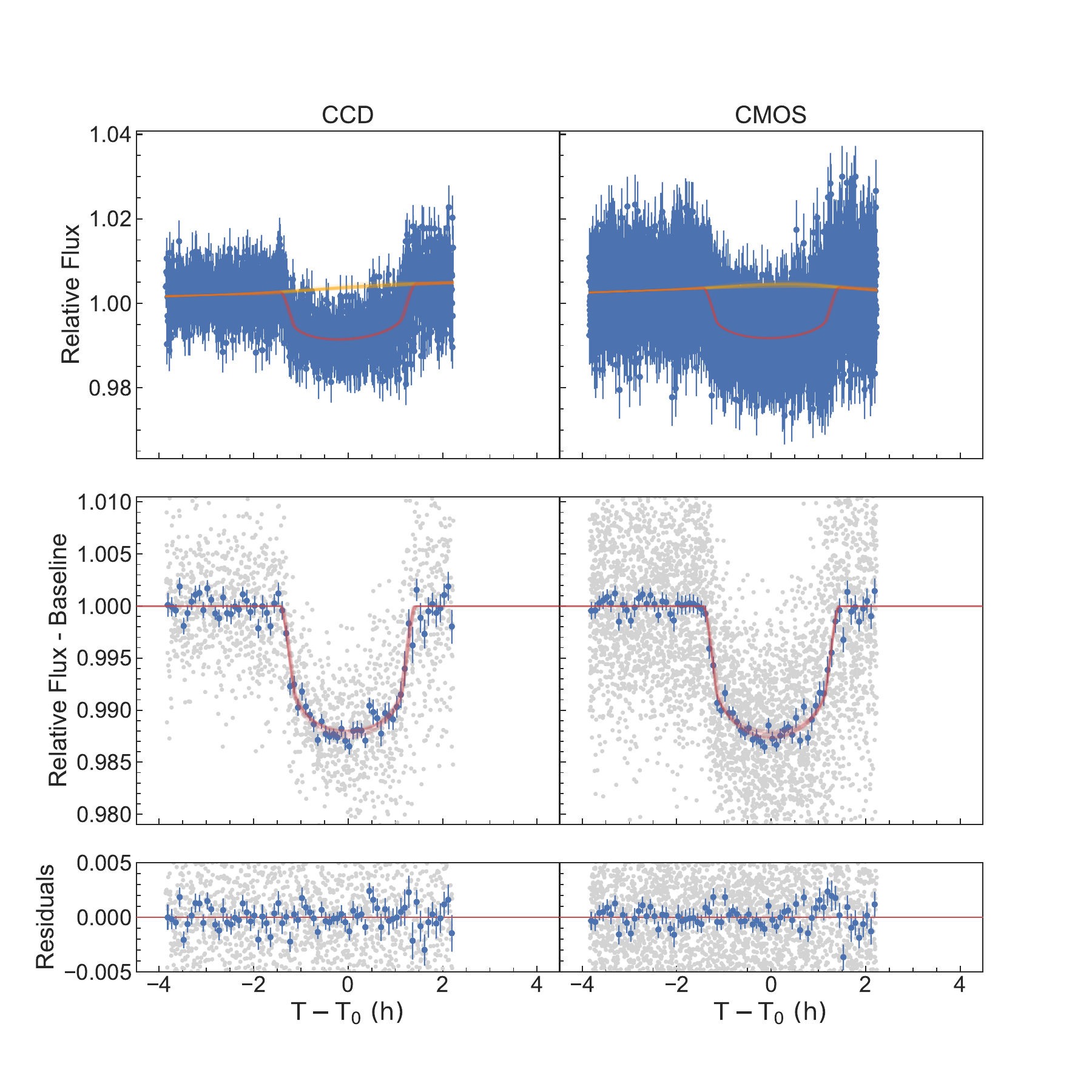}
\caption{Same as in \autoref{fig:WASP-4}. Normalized lightcurves for the transiting system of WASP-95 (TIC 144065872) from NGTS on 2024 August 07.}
\end{figure*}

\begin{table*}
	\centering
	\caption{\texttt{allesfitter} model priors and fitted values. Planetary System properties for the transiting system WASP-95 (TIC 144065872) from NGTS on 2024 August 07.}
    \begin{tabular}{lcccc}
    \hline
	\textbf{Parameter}&\textbf{Initial Guess}&\textbf{Prior}&\textbf{Fitted value CMOS}&\textbf{Fitted value CCD} \\
    \hline
    $R_p / R_{\star}$            & \FITNGrb          & $\mathcal{N}\left(0.1025,0.0005\right)$         & \CMOSNGrb	    & \CCDNGrb   \\
	$(R_\star + R_b) / a_b$      & \FITNGrsuma       & $\mathcal{N}\left(0.172266,0.005922\right)$     & \CMOSNGrsuma	& \CCDNGrsuma   \\
    $\cos{i_b}$                  & \FITNGinc         & $\mathcal{N}\left(0.027922,0.036611\right)$     & \CMOSNGinc	    & \CCDNGinc   \\
    Tc (BJD)		             & \FITNGtc	         & $\mathcal{N}\left(2458553.711158,0.000088\right)$ & \CMOSNGtc	    & \CCDNGtc	\\
    Orbital Period (days)	  	 & \FITNGperiod      & $\mathcal{N}\left(2.1846730,0.0000014\right)$   & \CMOSNGperiod	& \CCDNGperiod	\\
	$q_1$                        & \FITNGqone        & $\mathcal{N}\left(0.357,0.012\right)$           & \CMOSNGqone	& \CCDNGqone   \\
	$q_2$                        & \FITNGqtwo        & $\mathcal{N}\left(0.393,0.089\right)$           & \CMOSNGqtwo	& \CCDNGqtwo   \\
    \hline
    \end{tabular}
    \label{tab:planetary-WASP-95_0807}
\end{table*}

\begin{figure*}
\centering
\includegraphics[width=\textwidth]{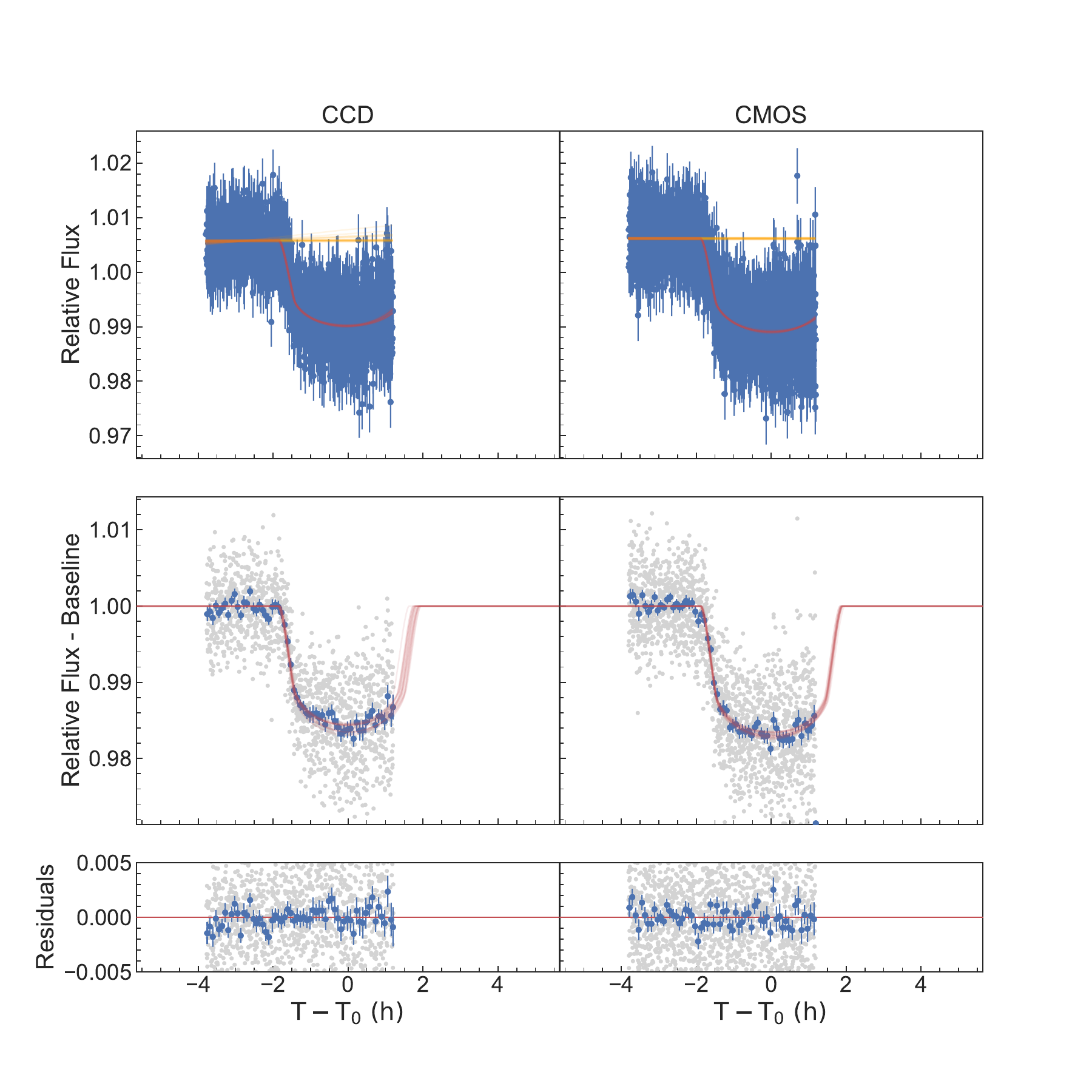}     
\caption{Same as in \autoref{fig:WASP-4}. Normalized lightcurves for the transiting system of KELT-10 (TIC 269217040) from NGTS on 2024 August 26.}
\end{figure*}

\begin{table*}
	\centering
	\caption{\texttt{allesfitter} model priors and fitted values. Planetary System properties for the transiting system KELT-10 (TIC 269217040) from NGTS on 2024 August 26.}
    \begin{tabular}{lcccc}
    \hline
	\textbf{Parameter}&\textbf{Initial Guess}&\textbf{Prior}&\textbf{Fitted value CMOS}&\textbf{Fitted value CCD} \\
    \hline
    $R_p / R_{\star}$            & \FITNHrb          & $\mathcal{N}\left(0.1190,0.0014\right)$         & \CMOSNHrb	    & \CCDNHrb   \\
	$(R_\star + R_b) / a_b$      & \FITNHrsuma       & $\mathcal{N}\left(0.119807,0.004107\right)$     & \CMOSNHrsuma	& \CCDNHrsuma   \\
    $\cos{i_b}$                  & \FITNHinc         & $\mathcal{N}\left(0.024258,0.015002\right)$     & \CMOSNHinc	    & \CCDNHinc   \\
    Tc (BJD)		             & \FITNHtc	         & $\mathcal{N}\left(2457612.49947,0.00041\right)$ & \CMOSNHtc	    & \CCDNHtc	\\
    Orbital Period (days)	  	 & \FITNHperiod      & $\mathcal{N}\left(4.1662541,0.0000016\right)$   & \CMOSNHperiod	& \CCDNHperiod	\\
	$q_1$                        & \FITNHqone        & $\mathcal{N}\left(0.350,0.010\right)$           & \CMOSNHqone	& \CCDNHqone   \\
	$q_2$                        & \FITNHqtwo        & $\mathcal{N}\left(0.385,0.081\right)$           & \CMOSNHqtwo	& \CCDNHqtwo   \\
    \hline
    \end{tabular}
    \label{tab:planetary-KELT-10_0826}
\end{table*}

\begin{figure*}
\centering
\includegraphics[width=\textwidth]{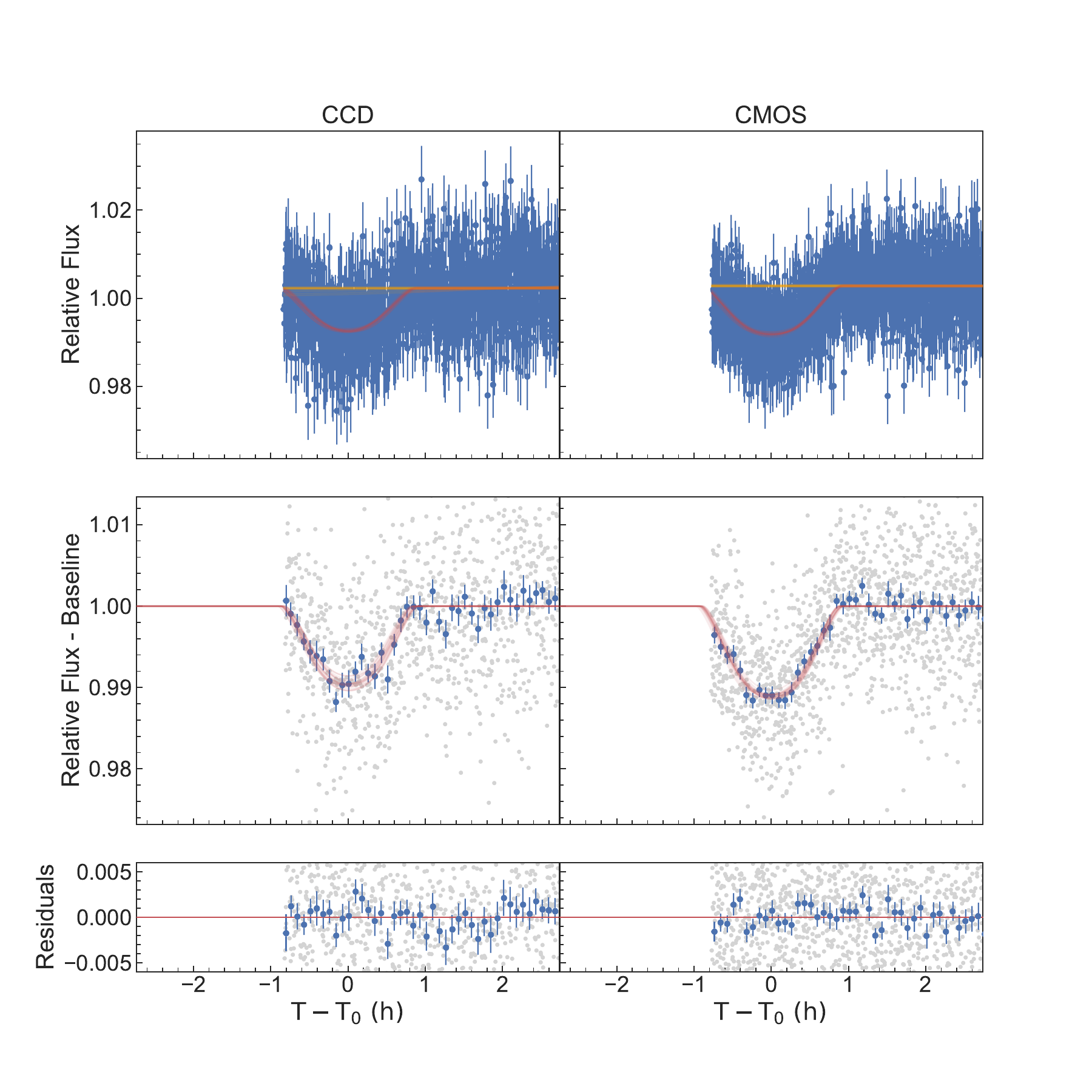}
\caption{Same as in \autoref{fig:WASP-4}. Normalized lightcurves for the transiting system of TOI-4463 (TIC 8599009) from NGTS on 2024 August 27.}
\end{figure*}

\begin{table*}
	\centering
	\caption{\texttt{allesfitter} model priors and fitted values. Planetary System properties for the transiting system TOI-4463 (TIC 8599009) from NGTS on 2024 August 27.}
    \begin{tabular}{lcccc}
    \hline
	\textbf{Parameter}&\textbf{Initial Guess}&\textbf{Prior}&\textbf{Fitted value CMOS}&\textbf{Fitted value CCD} \\
    \hline
    $R_p / R_{\star}$            & \FITNIrb          & $\mathcal{N}\left(0.11449,0.00397\right)$         & \CMOSNIrb	    & \CCDNIrb   \\
	$(R_\star + R_b) / a_b$      & \FITNIrsuma       & $\mathcal{N}\left(0.136414,0.004037\right)$     & \CMOSNIrsuma	& \CCDNIrsuma   \\
    $\cos{i_b}$                  & \FITNIinc         & $\mathcal{N}\left(0.107652,0.004857\right)$     & \CMOSNIinc	    & \CCDNIinc   \\
    Tc (BJD)		             & \FITNItc	         & $\mathcal{N}\left(2459291.64004,0.00032\right)$ & \CMOSNItc	    & \CCDNItc	\\
    Orbital Period (days)	  	 & \FITNIperiod      & $\mathcal{N}\left(2.8807198,0.0000028\right)$   & \CMOSNIperiod	& \CCDNIperiod	\\
	$q_1$                        & \FITNIqone        & $\mathcal{N}\left(0.372,0.007\right)$           & \CMOSNIqone	& \CCDNIqone   \\
	$q_2$                        & \FITNIqtwo        & $\mathcal{N}\left(0.400,0.071\right)$           & \CMOSNIqtwo	& \CCDNIqtwo   \\
    \hline
    \end{tabular}
    \label{tab:planetary-TOI-4463_0827}
\end{table*}

\begin{figure*}
\centering
\includegraphics[width=\textwidth]{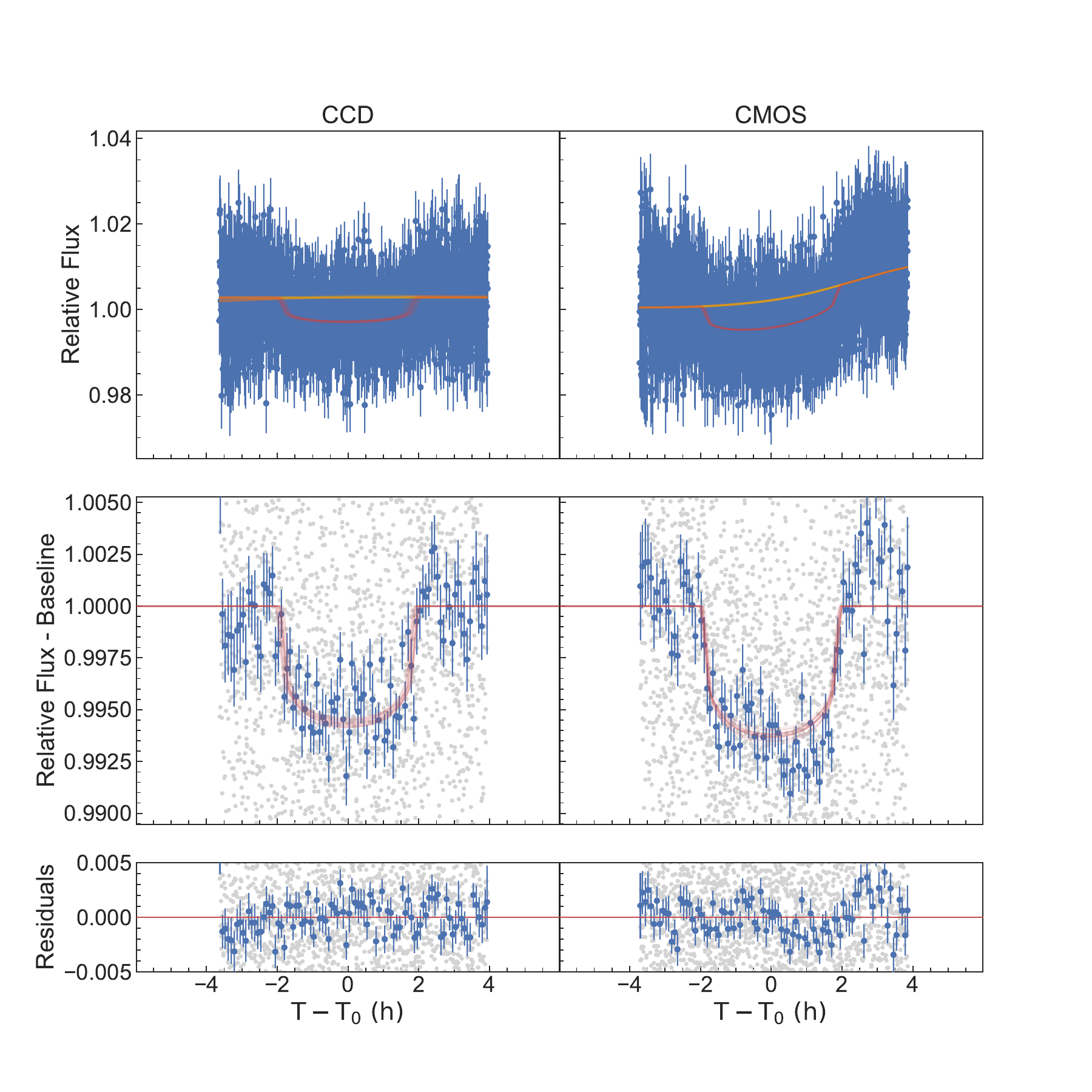}
\caption{Same as in \autoref{fig:WASP-4}. Normalized lightcurves for the transiting system of WASP-30 (TIC 9725627) from NGTS on 2024 August 28.}
\label{fig:wasp-30}
\end{figure*}

\begin{table*}
	\centering
	\caption{\texttt{allesfitter} model priors and fitted values. Planetary System properties for the transiting system WASP-30 (TIC 9725627) from NGTS on 2024 August 28.}
    \begin{tabular}{lcccc}
    \hline
	\textbf{Parameter}&\textbf{Initial Guess}&\textbf{Prior}&\textbf{Fitted value CMOS}&\textbf{Fitted value CCD} \\
    \hline
    $R_p / R_{\star}$            & \FITNKrb          & $\mathcal{N}\left(0.0705,0.0011\right)$         & \CMOSNKrb	    & \CCDNKrb   \\
	$(R_\star + R_b) / a_b$      & \FITNKrsuma       & $\mathcal{N}\left(0.121078,0.001989\right)$     & \CMOSNKrsuma	& \CCDNKrsuma   \\
    $\cos{i_b}$                  & \FITNKinc         & $\mathcal{N}\left(0.007505,0.008202\right)$     & \CMOSNKinc	    & \CCDNKinc   \\
    Tc (BJD)		             & \FITNKtc	         & $\mathcal{N}\left(2460551.738836,0.0068717\right)$ & \CMOSNKtc	    & \CCDNKtc	\\
    Orbital Period (days)	  	 & \FITNKperiod      & $\mathcal{N}\left(4.156736,0.00000202\right)$   & \CMOSNKperiod	& \CCDNKperiod	\\
	$q_1$                        & \FITNKqone        & $\mathcal{N}\left(0.330,0.006\right)$           & \CMOSNKqone	& \CCDNKqone   \\
	$q_2$                        & \FITNKqtwo        & $\mathcal{N}\left(0.373,0.065\right)$           & \CMOSNKqtwo	& \CCDNKqtwo   \\
    \hline
    \end{tabular}
    \label{tab:planetary-wasp-30}
\end{table*}

\begin{table}
\centering
\caption{Percentage ratio of transition pixels within star apertures}
\label{tab:transition}
\begin{tabular}{cccc}
\hline 
\hline
\textbf{\tess\ Magnitude bin}  & \textbf{ Stars affected by transition pixels}&     \textbf{Average number of pixels in transition \#/78}             \\ \hline  
6-7     &   2/2 &   3.50     \\
7-8    &   9/9 &   2.33     \\
8-9    &   10/10 &   2.40     \\
9-10    &   15/18 &   2.00     \\
10-11    &   53/57 &   1.81     \\
11-12    &   92/115 &   1.54     \\
12-13    &   85/196 &   0.65     \\   
13-14    &    1/378&   0.00     \\ 
14-15    &    1/547&   0.01     \\ 
15-16    &    6/942&   0.01     \\ \hline 
\end{tabular}
\end{table}





\bsp	
\label{lastpage}
\end{document}